\documentclass[aps,prl,preprint,groupedaddress,nofootinbib,preprintnumbers]{revtex4}
\usepackage{amsmath,amssymb,graphicx}
\newcommand{\bear}{\begin{eqnarray}  }
\newcommand{\eear}{\end{eqnarray}  }
\newcommand{\be}{\begin{equation} }
\newcommand{\ee}{\end{equation}}
\begin{document}
\preprint{\parbox[b]{1in}{ \hbox{\tt PNUTP-10/A01} \hbox{\tt IC/2010/007}}}

\title{Holographic aspects of three dimensional  QCD from string theory}

\author{Deog Ki Hong}
\email[]{dkhong@pusan.ac.kr}
\affiliation{Department of
Physics,   Pusan National University,
             Busan 609-735, Korea}
\author{Ho-Ung Yee}
\email[]{hyee@ictp.it}
\affiliation{ICTP, High Energy, Cosmology and Astroparticle Physics,
Strada Costiera 11, 34151, Trieste, Italy}


\date{\today}

\begin{abstract}
We study two aspects of 3D QCD with massless fermions in a holographic set-up from string theory, based on D3/D7 branes; parity anomaly and baryons as baby Skyrmions.
We first give a novel account of parity anomaly of 3D QCD with odd number  of flavors from the IR holographic viewpoint by
observing a subtle point in D7 brane embeddings with a given fixed UV theory.
We also discuss its UV origin in terms of weakly coupled D-brane pictures.
We then focus on the parity-symmetric case of even number of $N_F$ flavors, and study baryons
in the holographic model. We identify the monopoles of $U(N_F)$ gauge theory dynamically broken down to $U({N_F\over 2})\times U({N_F\over 2})$ in the holographic 4 dimensional bulk as a holographic counter-part of 3D baby-Skyrmions for baryons in large N limit, and work out some details how the mapping goes. In particular, we show that the correct baryon charges emerge from the Witten effect with a space-varying $\Theta$ angle.

\end{abstract}


\maketitle

\section{Introduction}
\label{intro}

Recent study in AdS/CFT correspondence has shown that certain gauge theories are dual to gravity theories in a limit, where the number of fields is very large, $N_c\to\infty$, and also their coupling is very strong, $\lambda =g_{YM}^2 N_c \to\infty$ \cite{Maldacena:1997re,Witten:1998qj}. In such limit the dual gravity  is weakly coupled and also its quantum effects are largely suppressed, which makes  the theory rather easily solvable.  This strong and weak coupling duality is not only interesting but also has far-reaching consequences, as it offers new understanding of strongly interacting systems.

Though the duality between gauge and gravity theories is rigorously established only for specific theories such as  ${\cal N}=4$ super Yang-Mills theory,
it has been widely used in recent years to construct  models for more traditional theories like QCD or strongly correlated systems, which are otherwise very difficult to solve \cite{Aharony:1999ti}. Those gravity dual models are often called holographic models, since at low energy the gravity dual is defined in one higher dimension than that of gauge theory.

As an attempt to construct the gravity dual of QCD with no flavors, string theory with a stack of $N_{c}$ D4 branes in  $R^{4}\times S^{1}$ is studied as a low-energy effective theory of pure glue theory for the large number of colors, $N_{c}\gg1$, in the strong 't Hooft coupling  limit, $\lambda (\equiv g_{\rm YM}^{2}N_{c})\gg1$ \cite{Witten:1998zw}.  The dual theory exhibits color confinement and has color-singlet excitations, corresponding to  glueballs, whose spectrum agrees well with lattice results, after matching  the gravity scale  with  the lattice data \cite{Csaki:1998qr}. Chiral quarks are later introduced in the gravity dual by Sakai and Sugimoto as open strings between D4 and D8 branes \cite{Sakai:2004cn,Sakai:2005yt}. Treating the D8 brane as a probe in the background of D4 branes, they find the spontaneous chiral symmetry breaking is realized geometrically in the D4 background. The low energy excitations of open strings attached to D8 branes include the  whole tower of vector mesons as well as (massless) pions. The spectrum and couplings of mesons are found to agree reasonably well with experimental data. See refs.\cite{Kruczenski:2003be,Kruczenski:2003uq,Babington:2003vm} for similar constructions using D3/D7 branes. Furthermore the holographic model  offers natural explanation for the phenomenological rules, found empirically in hadron physics, such as  vector meson dominance \cite{Hong:2004sa,Hong:2007dq}.

In this paper we study three dimensional quantum chromodynamics, ${\rm QCD}_{3}$ using a similar holographic dual construction. The construction was first introduced by Rey \cite{rey,Rey:2008zz}, where he first discussed relevant symmetry-breaking patterns, and we intend to study two interesting aspects of 3D QCD in the holographic set-up: parity anomaly \cite{Redlich:1983kn,Redlich:1983dv,Niemi:1983rq} and holographic baby Skyrmions as baryons of 3D QCD. Three dimensional gauge theories have been intensively studied as toy models to understand non-pertubative properties of QCD, as they exhibit confinement as well as dynamical symmetry breaking. Confinement is for instance recently established in three dimensional Yang-Mills theory~\cite{Karabali:1997wk}.  Three dimensional gauge theories arise in the high temperature limit of four dimensional gauge theories and also describe  the low energy physics of planar condensed matter systems, exhibiting dynamical generation of mass. As a gravity dual of  ${\rm QCD}_{3}$ with $N_{F}$ flavors one can introduce  $N_{F}$ number of D7 branes in the background of a stack of $N_{c}$ D3 branes  in $R^{3}\times S^{1}$ with anti-periodic boundary condition for fermions.
The low energy excitations of the gravity dual are color singlets and the constituent quarks get dynamical masses. We provide a novel explanation of parity anomaly with odd number of flavors in this holographic set-up. We then show that the baryons of ${\rm QCD}_{3}$ are realized as 4-dimensional monopoles in the gravity dual, which are holographic version of magnetic vortices, or baby Skyrmions.

 The plan of our paper is following.  We first review the ${\rm QCD}_{3}$ and discuss how its symmetries are realized. Then, we move on to the gravity dual model of ${\rm QCD}_{3}$, and discuss parity anomaly in the holographic set-up. Finally we discuss how baryons are identified, focusing on the derivation of their correct charges under the flavor symmetry.

\section{Review on field theory perspectives}

\subsection{ Parity anomaly in three dimensions}
\label{anomaly}
In three dimensional spacetime spin-1/2 fermions are described by two-component spinor, the minimal spinor representation of SO(2,1). A free massive fermion is described by a Lagrangian density
\begin{equation}
{\cal L}_{\rm free}=\bar\psi\left(i\gamma_{2\times2}^{\mu}\partial_{\mu}-m\right)\psi,\label{free}
\end{equation}
where $\mu=0,1,2$ and the ${2\times2}$ Dirac matrices are given as
\begin{equation}
\gamma_{2\times2}^{0}=\begin{pmatrix} 1 & 0 \\ 0 & -1 \end{pmatrix},\quad
\gamma_{2\times2}^{1}=\begin{pmatrix} 0 & i \\ i & 0\end{pmatrix},\quad
\gamma_{2\times2}^{2}=\begin{pmatrix} 0 & 1 \\ -1 & 0 \end{pmatrix}\,,
\end{equation}
which satisfy the Clifford algebra, $\{\gamma_{2\times2}^{\mu},\gamma_{2\times2}^{\nu}\}=2\,\eta^{\mu\nu}$ in (2+1) dimensions.

The free fermion Lagrangian (\ref{free}) is invariant under the ${\rm U}(1)$ fermion number symmetry,
\begin{equation}
\psi(x)\mapsto \psi^{\prime}(x)=e^{i\theta}\psi(x), \quad \bar\psi(x)\mapsto\bar\psi^{\prime}(x)=e^{-i\theta}\bar\psi(x)\,.
\end{equation}
Unlike in even dimensions the mass parameter $m$  is real in odd dimensions, defined up to a sign, and we do not have chiral symmetry in the massless limit $m=0$, since there is no $\gamma_{5}$ that anticommutes with all $\gamma$ matrices.  When $m=0$, however, we have an enhanced discrete symmetry,  the parity invariance,  at the classical level, which is broken though by  quantum effects \cite{Redlich:1983kn,Redlich:1983dv,Niemi:1983rq} .

Under the parity, $P_{2}$, $x=(t,x_{1},x_{2})\mapsto x^{\prime}=(t,-x_{1},x_{2})$, the fermions transform as
\begin{equation}
\psi(x)\underset{P_{2}}\longmapsto\psi^{\prime}(x^{\prime})=e^{i\delta}\gamma_{2\times2}^{1}\psi(x)\,,\quad
\bar\psi(x)\underset{P_{2}}\longmapsto\bar\psi^{\prime}(x^{\prime})=e^{-i\delta}\bar\psi(x)\gamma_{2\times2}^{1}\,,
\end{equation}
where $\delta$ is an arbitrary real phase. The mass term in the free Lagrangian (\ref{free}) changes its sign under the parity,
\begin{equation}
{\cal L}_{\rm free}=\bar\psi\left(i\gamma_{2\times2}^{\mu}\partial_{\mu}-m\right)\psi
=\bar\psi^{\prime}(x^{\prime})\left(i\gamma_{2\times2}^{\mu}\partial^{\prime}_{\mu}+m\right)\psi^{\prime}(x^{\prime})\,.
\end{equation}
The Lagrangian is therefore invariant under parity, if $m=0$. However the parity is broken at the quantum level, once an interaction is introduced for fermions.

\begin{figure}[t]
	\centering
	\includegraphics[width=0.5\textwidth]{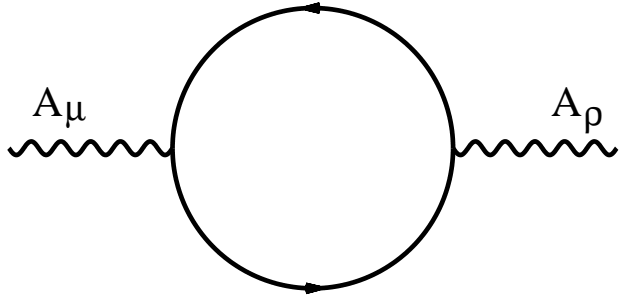}%
		\caption{\label{parity} One-loop diagram of planar fermions generating parity anomaly. }
\end{figure}
Suppose the fermions couple to external fields $A_{\mu}$~\footnote{In three dimensions the spin is just an additive quantum number, since the little group of the Lorentz symmetry is $U(1)$. Hence, the external fields $A_{\mu}$ could be spin-zero fields or spin-1 fields.}
 with a strength $e$  as
\begin{equation}
{\cal L}_{\rm int}=-eA_{\mu}j^{\mu}\,,
\end{equation}
 where $j^{\mu}=\bar\psi(x)\gamma^{\mu}\psi(x)$ is the fermion current.  When expanded in powers of derivative, the two-point function of the external fields contains at low energy a Chern-Simons term, breaking the parity,
  \begin{equation}
{\cal L}_{\rm CS}=\frac{e^{2}}{8\pi}\frac{m}{|m|}\epsilon^{\mu\nu\rho}A_{\mu}\partial_{\nu}A_{\rho}\,,
\end{equation}
which arises at one-loop shown in Fig.~\ref{parity}\,.  (Here $m$ is either the fermion mass or the Pauli-Villars regulator in the case of massless fermions, introduced to regulate the ultraviolet divergence of the one-loop diagram.) In fact, to regularize divergence one needs to introduce a gauge invariant Pauli-Villars fermion which contributes a {\it half} of the above result. The other half comes from expanding finite non-local effective action in powers of $1/m$ in a large mass limit. For massless case or small mass compared to the momentum scale of interests, one in general cannot expand the non-local effective action to get the above full expression, while only a half integral Chern-Simons term from the Pauli-Villars fermion is present explicitly. The two choices of Pauli-Villars mass sign would therefore result in two different theories whose effective actions differ by a {\it single} unit of Chern-Simons term \cite{Redlich:1983kn}.

If the number of fermion flavors is even, we can cancel the parity anomaly by choosing an opposite sign for mass or for the regulator of each flavor pair $(\psi_{i},\psi_{i+N})$, since the fermion mass is defined up to a sign, as
\begin{equation}
{\cal L}=\sum_{i=1}^{n}\left[\bar\psi_{i}\left(i\gamma_{2\times2}^{\mu}\partial_{\mu}-m_{i}\right)\psi_{i}+
\bar\psi_{i+n}\left(i\gamma_{2\times2}^{\mu}\partial_{\mu}+m_{i}\right)\psi_{i+n}
\right]+{\cal L}_{\rm int}\,.
\label{even}
\end{equation}
Equivalently, if we introduce four-component spinors (reducible spinors) and $4\times4$ $\gamma$ matrices,
\begin{equation}
\Psi_{i}=\begin{pmatrix} \psi_{i}\\ \psi_{i+n} \end{pmatrix},\qquad
\gamma^{\mu}=\gamma^{\mu}_{2\times2}\otimes\begin{pmatrix} 1 & 0 \\ 0 &-1 \end{pmatrix}\,,
\label{four}
\end{equation}
the mass term for the $i$-th pair becomes
\begin{equation}
m_{i}\bar\psi_{i}\psi_{i}-m_{i}\bar\psi_{i+n}\psi_{i+n}=m_{i}\bar\Psi_{i}\Psi_{i}\,.\label{pair-mass}
\end{equation}
By combining a discrete part ($Z_{2}$) of flavor symmetries, we may redefine parity~\footnote{$P_{4}$ is what we get from the parity of four-dimensional theories when dimensionally reduced.}, $P_{4}$,
under which $x=(t,x_{1},x_{2})\mapsto x^{\prime}=(t,-x_{1},x_{2})$ and for each pair of flavors
\begin{equation}
\psi_{i}(x)\underset{P_{4}}\longmapsto e^{i\delta}\gamma_{2\times2}^{1}\psi_{i+n}(x)\,;\quad
\psi_{i+n}(x)\underset{P_{4}}\longmapsto e^{i\delta}\gamma_{2\times2}^{1}\psi_{i}(x)
\end{equation}
Or in terms of four-component spinors
\begin{equation}
\Psi_{i}(x)\underset{P_{4}}\longmapsto\Psi^{\prime}_{i}(x^{\prime})=e^{i\delta}
\begin{pmatrix} 0 & \gamma_{2\times2}^{1} \\ \gamma_{2\times2}^{1} &0 \end{pmatrix}
\Psi_{i}(x)\,.
\end{equation}
The mass term (\ref{pair-mass}) is then $P_{4}$ invariant and is called a parity-invirant mass since the mass term does not generate Chern-Simons terms for the (3D) gluons in the effective action, $S_{\rm eff}(A)=\ln\det\left(i\!\!\not\!\partial+e\!\!\not\!A\right)$, when it  is gauge-invariantly regularized.

Now, if we take $m_{i}\to0$,
we have enhanced  ${\rm U}(2)$ ``chiral symmetry'', generated by
\begin{equation}
{\bf 1}_{4\times4},\quad \gamma^{3}=i\begin{pmatrix} 0 & {\bf 1}_{2\times2} \\ {\bf 1}_{2\times2} &0 \end{pmatrix}, \quad
\gamma^{5}=i\begin{pmatrix} 0 & {\bf 1}_{2\times2} \\ -{\bf 1}_{2\times2} &0 \end{pmatrix}, \quad {\rm and}\quad
\frac12\left[\gamma^{3},\gamma^{5}\right]\,,
\end{equation}
where ${\bf 1}$ denotes identity matrices. For $2n$ massless flavors we have then ${\rm U}(2n)$ ``chiral'' symmetry. 

\subsection{ Dynamical Mass Generation in  three dimensional QCD}
\label{mass}
In strongly interacting systems the ground states often have less symmetries than Hamiltonian due to strong dynamics.
Coleman and Witten~\cite{Coleman:1980mx} have shown, with reasonable assumptions, that
the chiral symmetry of four dimensional QCD  breaks down spontaneously in the limit of a large number of colors by
the 't Hooft anomaly matching condition for chiral currents together with the vacuum energy analysis. In three dimensional QCD, however, there is no axial anomaly to constrain the spectrum at low energy, though all the assumptions made in Ref.~\cite{Coleman:1980mx} go in parallel for three dimensional QCD, including confinement~\cite{Karabali:1997wk}. But, fortunately, Vafa and Witten~\cite{Vafa:1983tf} found that the positivity of path integral measure of vector-like gauge theories restricts the pattern of symmetry breaking, which is applicable to theories in any dimension. For instance, they showed, among others,  that parity and also vector symmetries can not be broken spontaneously in vector-like gauge theories.
By this Vafa-Witten theorem, one can show that the ${\rm U}(2n)$ ``chiral''  symmetry of three dimensional QCD with $2n$ massless quarks has to break down dynamically to ${\rm U}(n)\times{\rm U}(n)$, if spontaneously broken~\cite{Vafa:1984xh}.

If we introduce a small bare mass for quarks as $m\left(\bar\psi_{i}\psi_{i}-\bar\psi_{i+n}\psi_{i+n}\right)$ or $m\bar\Psi_{i}\Psi_{i}$ for all $i=1,\cdots n$ with $m\to0$,
the quark determinant becomes in the Euclidean space
\begin{equation}
\lim_{m\to0}\det\left[(i\!\not\!\!D+im)(i\!\not\!\!D-im)\right]^{n}=\lim_{m\to0}\det\left[-\left(\not\!\!D\right)^{2}+m^{2}\right]^{n}
\end{equation}
which gives a positive measure for three dimensional QCD in the Euclidean functional integration. Therefore the parity ($P_{4}$) can not be broken spontaneously according to the Vafa-Witten theorem~\cite{Vafa:1984xh}.

Whether the ``chiral"  ${\rm U}(2n)$ symmetry is spontaneously broken or not is a dynamical issue. Indeed, it has been shown by the Schwinger-Dyson analysis there is a phase in three dimensional gauge theories where  the ``chiral" ${\rm U}(2n)$ symmetry is spontaneously broken down to ${\rm U}(n)\times{\rm U}(n)$, generating dynamical mass for fermions~\cite{Pisarski:1984dj,Appelquist:1986qw,Appelquist:1989tc,Hong:1993qk}. (We will see in the later sections that the spontaneous breaking of ${\rm U}(2n)$  down to  ${\rm U}(n)\times{\rm U}(n)$ is geometrically realized in the D brane picture.)

In the large $N_{c}$ limit we can use an argument similar to that of Coleman and Witten~\cite{Coleman:1980mx} to arrive at the same conclusion as above for the symmetry breaking.
Suppose the order parameter for the symmetry breaking is a quark bilinear,
\begin{equation}
M_{i}^{j}=\left<\bar\psi_{i}\psi^{j}\right>,
\end{equation}
which is a Hermitian matrix and $i,j=1,\cdots,2n$. The order parameter transforms under the ``chiral''
${\rm U}(2n)$ as
\begin{equation}M\longmapsto g^{\dagger}Mg,\qquad g\in{\rm U}(2n)\,
\end{equation}
and under the $P_{4}$ parity
\begin{equation}
M\underset{P_{4}}\longmapsto P_{4}^{-1}MP_{4}=-I_1 M I_1\,,
\end{equation}
where
\be
I_{1}=\begin{pmatrix} 0 & {\bf 1}_{n\times n}  \\ {\bf 1}_{n\times n}&0 \end{pmatrix}\quad.
\ee

If we expand the effective potential $V$ for the order parameter,
it contains ${\rm Tr}\,M^{r}$, ${\rm Tr}\,M^{r}\times {\rm Tr}M^{s}$, {\it etc}, but only even powers of $M$ because of the $P_{4}$-parity invariance.
Since the trace arises from the quark loops and the single quark loops are dominant in the large $N_{c}$,
we have, to leading order in $1/N_{c}$,
\begin{equation}
V=N_{c}\,{\rm Tr}\,F(M^{2})=N_{c}\sum_{i}F(\lambda_{i})\,,
\end{equation}
where $F$ is some $N_{c}$-independent function and $\lambda_{i}$ ($i=1,\dots,2n$) are the eigenvalues of $M^{2}$. Since the eigenvalues are independent variables, the minimum of the effective potential occurs when each eigenvalues are at the minimum of $F$. If the minimum occurs at a nonzero value, $\lambda_{i}=\kappa^{2}\,(\ne0)$, the eigenvalues of the order parameter take either $+\kappa$ or $-\kappa$, having ${2n}$-fold degeneracy. But, the parity ($P_{4}$) conservation picks a unique ground state out of $2n$-fold degeneracy, that gives ${\rm Tr}\,M=0$ and preserves ${\rm U}(n)\times{\rm U}(n)$.

\subsection{ Bosonization  of  three dimensional QCD}
\label{bosonization}
When the `chiral' symmetry is broken in three dimensional QCD by vacuum condensates, the relevant degrees of freedom at low energy are Nambu-Goldstone bosons. Consider composite fields
\begin{equation}
\phi(x)=\lim_{y\to x}\frac{|x-y|^{\gamma}}{\kappa}\,\psi(y)\bar\psi(x)\,,
\end{equation}
which transform under the ``chiral'' symmetry, $g\in {\rm U}(2n)$  as
\begin{equation}
\phi(x)\longmapsto g\phi g^{\dagger}\,.
\end{equation}
Since in the ground state we have
\begin{equation}
\left<\phi\right>=I_{3},\qquad {\rm with}\quad I_{3}=\begin{pmatrix} {\bf 1}_{n\times n} &0 \\ 0&-{\bf 1}_{n\times n}  \end{pmatrix},
\end{equation}
the Nambu-Goldstone bosons are described by
\begin{equation}
\phi(x)=\,g(x)I_{3}g^{\dagger}(x), \quad g\in {\rm SU}(2n)\,,
\label{ng}
\end{equation}
which satisfies the constraint ${\rm Tr}\left(\phi^{2}\right)=2n$.
The Lagrangian density for the Nambu-Goldstone bosons are given at the low energy as
\begin{eqnarray}
{\cal L}_{\rm B}=\frac{f_{\pi}^{2}}{2}\,{\rm Tr}\,\partial_{\mu}\phi\,\partial^{\mu}\phi
                        =f_{\pi}^{2}\,{\rm Tr}\left(\partial_{\mu}g\,\partial^{\mu}g^{\dagger}-j_{\mu}j^{\mu}\right)\,,
\end{eqnarray}
where  $j_{\mu}=\frac{1}{2i}\left[g^{\dagger}\partial_{\mu}gI_{3}-\left(\partial_{\mu}g^{\dagger}\right)gI_{3}\right]$.

Introducing auxiliary fields $\bar A_{\mu}$ and rescaling $g\mapsto g/f_{\pi}$, and then adding a constant term ${\rm Tr}\left(j_{\mu}-{\bar A_{\mu}}\right)^{2}$, we rewrite the low-energy Nambu-Goldstone Lagrangian as
\begin{equation}
{\cal L}_{B}
={\rm Tr}\,\left[\left(\partial_{\mu}-i\bar A_{\mu}\right)g^{\dagger}\left(\partial_{\mu}+i\bar A_{\mu}\right)g\right]\,.
\end{equation}
Describing the Nambu-Goldstone boson fields $\phi(x)$ in terms of $g(x)$, we have introduced too many degrees of freedom, since
the transformation $g\mapsto u^{\dagger}gu$ leaves $\phi(x)$ invariant, if $u\in {\rm SU}(n)_{1}\times {\rm SU}(n)_{2}\times {\rm U}(1)_{3}$, the unbroken flavor symmetry, where ${\rm U}(1)_{3}$ is an Abelian group generated by $I_{3}$. This redundancy in the fields can be removed by gauging the unbroken symmetry under which the auxiliary fields
\begin{equation}
\bar A_{\mu}\longmapsto u^{\dagger}\bar A_{\mu}u-i\partial_{\mu}u^{\dagger}\,,
\end{equation}
where $\bar A_{\mu}=\bar A_{1\mu}\oplus\bar A_{2\mu}\oplus\bar A_{3\mu}$. The gauge fields $A_{1}$, $A_{2}$, and $A_{3}$ are those of ${\rm SU}(n)_{1}$, ${\rm SU}(n)_{2}$, and ${\rm U}(1)_{3}$, respectively.

As shown in Ref.~\cite{Ferretti:1992fga,Ferretti:1992fd}, the low energy effective Lagrangian should contain Chern-Simons terms for the auxiliary fields because of  $P_2$-anomaly in the
 two-point functions of diagonal flavor currents~\footnote{The two-point functions of flavor currents off-diagonal in flavor space give similar results, but the calculation of the two-point functions of diagonal flavor currents is enough to find the symmetry breaking pattern.}
$j_{i}^{\mu}=\bar\psi_{i}\gamma^{\mu}_{2\times2}\psi_{i}$ ($i=1,\cdots,2n$).
The non-vanishing components at low energy $k\to0$ are
\begin{equation}
\left<j^{\mu}_{i}\left(k\right)j^{\nu}_{j}\left(-k\right)\right>=\lim_{m\to0}\frac{m_{i}}{|m_{i}|}\delta_{ij}\frac{N_{c}}{4\pi}\epsilon^{\mu\lambda\nu}\,k_{\lambda}\,,
\label{pattern}
\end{equation}
where $m_{i}=m$ for $i=1,\cdots, n$ and $m_{i}=-m$ for $i=n+1,\cdots, 2n$.
We see that  the two-point functions respect the $P_{4}$ parity as well as the vector-like flavor symmetry $U(n)\times U(n)$, and it does not preserve the full $U(2n)$ symmetry even in the limit $m\to0$. The two point functions (\ref{pattern}) give rise to Chern-Simons terms for auxiliary fields $\bar A_\mu$ that couple to flavor currents.

The low energy effective Lagrangian of ${\rm QCD}_{3}$ should therefore include Chern-Simons terms but such a way that preserves $P_{4}$ parity,
\begin{equation}
{\cal L}_{\rm eff}={\cal L}_{B}+\frac{k}{4\pi}\,{\cal L}_{\rm CS}\left({\bar A_{1}}\right)-\frac{k}{4\pi}\,{\cal L}_{\rm CS}\left(\bar A_{2}\right)+\cdots\,,
\label{eff}
\end{equation}
where the ellipsis denotes terms with higher derivatives and the Chern-Sinons terms are given as
${\cal L}_{\rm CS}(\bar A)={\rm Tr}\,(\bar A{\rm d}\bar A+\frac23{\bar A}^{3})$. The coefficient of the Chern-Simons term is fixed by matching the $P_{2}$ anomaly, Eq.~(\ref{pattern}), $k=N_{c}$.

Although the above two-point functions and the resulting (\ref{eff}) might seem to break $U(2n)$ explicitly, it has no dynamical consequences when we turn off external potential $\bar A_\mu$.
It can also be understood from the fact that Pauli-Villars fermions whose masses explicitly violate $U(2n)$ symmetry down to $U(n)\times U(n)$ enter only the two-point 1-loop diagram of Fig.~\ref{parity}, while higher-point diagrams are finite due to super-renormalizability. The only $U(2n)$-breaking effect from Pauli-Villars masses is therefore the local flavor Chern-Simons term (\ref{eff}) which vanishes for $\bar A_\mu=0$, and the conservation Ward identity of full $U(2n)$ symmetry is intact. We will observe the precisely same feature in the holographic set-up later.

The manifold of Nambu-Goldstone fields of three dimensional QCD has a nontrivial topology
\begin{equation}
\Pi_{2}\left(\frac{{\rm SU}(2n)}{{\rm SU}(n)\times{SU}(n)\times {\rm U}(1)_{3}}\right)=\Pi_{1}\left({\rm U}(1)_{3}\right)=Z\,.
\end{equation}
It should therefore allow a topological soliton or a vortex, similar to the baby Skyrmion, whose topological charge corresponds to the winding number in the ${\rm U}(1)_{3}$ space or its magnetic flux,
\begin{equation}
Q=\int{\rm d}^{2}x\,J_{0}=\frac{1}{2\pi}\int{\rm d}^{2}x\,\epsilon_{0ij}\partial_{i}\bar A_{3j}\,.
\end{equation}
Recall that because $I_3={\rm diag}(1_n,-1_n)$, the unit $U(1)_3$ vortex is nothing but a vortex of $U(1)\times U(1)\subset U(n)\times U(n)$ theory with charges $(+1,-1)$, which will be consistent with our holographic description later.

We now show that the winding number or the ${\rm U}(1)_{3}$ vortex number is nothing but the baryon number. To this end we consider another discrete anomaly in addition to the ones considered in Eq.~(\ref{pattern}), which arises in the two-point function of the quark number current and the chiral quark-number current at low energy,
\begin{equation}
\left<J^{\mu}\left(k\right)J^{\nu}_{35}\left(-k\right)\right>=\frac{N_{c}}{2\pi}\epsilon^{\mu\lambda\nu}\,k_{\lambda}+{\cal O}(k^{2})\,,
\label{discrete}
\end{equation}
where $J^{\mu}=\bar\Psi\gamma^{\mu}\Psi$ is the quark-number current and
$J^{\mu}_{35}=\bar\Psi\gamma^{\mu}\gamma^{3}\gamma^{5}\Psi$ is the chiral quark-number current. (Note here that we are using the four-component notations defined in Eq.~(\ref{four}).)
We promote the group-valued function $g(x)$ in Eq.~(\ref{ng}) to take values in ${\rm U}(2n)$ and then introduce a ${\rm U}(1)$ gauge fields $A_{\mu}$ to remove the redundancy. The low-energy effective Lagrangian (\ref{eff}) then should have a mutual Chern-Simons term to match the discrete anomaly,
\begin{equation}
{\cal L}_{\rm eff}\ni{\cal L}_{mCS}=\frac{N_{c}}{2\pi}\epsilon^{\mu\nu\lambda}A_{\mu}\partial_{\nu}\bar A_{3\lambda}
\end{equation}
and the gauge fields in Eq.~(\ref{ng}) contain ${\rm U}(1)$ gauge field $A_{\mu}$  or $\bar A_{\mu}\to \bar A_{\mu}+A_{\mu}$.

Since the ${\rm U}(1)$ gauge fields $A_{\mu}$ is associated with the quark number, the quark number current in the effective theory is
\begin{equation}
\left<J^{\mu}\right>=\frac{\delta S_{\rm eff}(A)}{\delta A_{\mu}}=\frac{N_{c}}{2\pi}\epsilon^{\mu\nu\lambda}\partial_{\nu}\bar A_{3\lambda}+\cdots\,,
\end{equation}
 where the ellipsis denote the quark currents from the phase of $g(x)$. We therefore see that the vortex of unit magnetic charge of $U(1)_3$, or equivalently the magnetic charge $(+1,-1)$ in $U(1)\times U(1)\subset U(n)\times U(n)$, carries $N_{c}$ quark number or unit baryon number.
We will reproduce this result in the holographic set-up, too.

\section{ Holographic Parity Anomaly of 3D QCD from String Theory}
\label{anomaly}

String theory can provide models of gauge theories by suitably engineering D-brane configurations
in the weak-coupling limit. By going over to a strongly coupled regime in a controlled way including back-reactions to the surrounding geometry, the resulting gravity theory may naturally be conjectured to be a holographic dual of the gauge theory  identified in the D-brane set-up. Consistency of the duality typically requires a large color and strong coupling limit.
Geometry/gravity itself maps to large N color dynamics, and one is often allowed to introduce `probe' branes in the geometry neglecting their back-reactions, which describes `quenched' dynamics of small number of additional flavor degrees of freedom \cite{Karch:2002sh}.
This philosophy can also been applied to 3 dimensional QCD with small number of Dirac fermions using D3/D7 branes. While
the geometry of large number of circle-compactified D3 branes has appeared and studied quite some time ago along the same line proposed by Witten for 4D Yang-Mills theory \cite{Aharony:1999ti,Csaki:1998qr}, introducing flavor D7 branes in ways that we will discuss
was first introduced in Ref.~\cite{rey}, and also discussed in Ref.~\cite{Davis:2008nv,Alanen:2009cn,Fujita:2009kw} in a different context\footnote{See also Ref.\cite{Leigh:2008tt} for a different gravity realization of parity symmetry breaking.}. Our discussion will contain a brief account of this well-known construction for completeness \footnote{Note that Ref.\cite{Davis:2008nv} also discusses parity anomaly from flavor symmetry in the NJL-type non-compact D3/D7 brane construction, and some of its contents overlaps with our next section. Our main result in this section is about parity anomaly of color interactions.}.

Being a super-renormalizable theory, 3D QCD with $N_F \ll N_c$ is expected to
be strongly coupled in the infrared regime. Because holographic models are generally said to be valid
in the strongly coupled regime, we expect this holographic model might capture correctly low energy physics of the 3D QCD. In turn, this also implies one shouldn't rely on the holographic model at stake in far UV regime, where typically holographic models have different UV completions from the gauge theory,  though they belong to the same universality in IR regime.

Since parity anomaly should be something that belongs to universal low energy phenomena, as every statement about anomalies does, it is logical to expect that one should be able to find its manifestation in the holographic model too, which will be shown to be  indeed so in the next subsection.

A set of D3 branes in flat space gives a conformal field theory with marginally tunable coupling constant,
and it is meaningful to have a weak-coupling limit to talk about D-branes. However, it seems more subtle to study  D-branes in non-conformal cases like our current situation. The weak coupling picture of D3/D7 branes should correspond to a different UV completion of the 3D QCD where the theory goes over to weakly coupled 4D ${\mathcal N}=4$ super Yang-Mills theory compactified on a circle with anti-periodic boundary condition for fermions plus ${\mathcal N}=2$ hypermultiplets. However, the universal character of parity anomaly should also be relevant here again, so it is logical to expect the anomaly to appear  even in this UV D-brane picture. We will discuss this in the subsection subsequent to the next subsection.

\subsection{Infrared strongly coupled picture}

\begin{figure}[t]
	\centering
	\includegraphics[width=10cm,height=9cm]{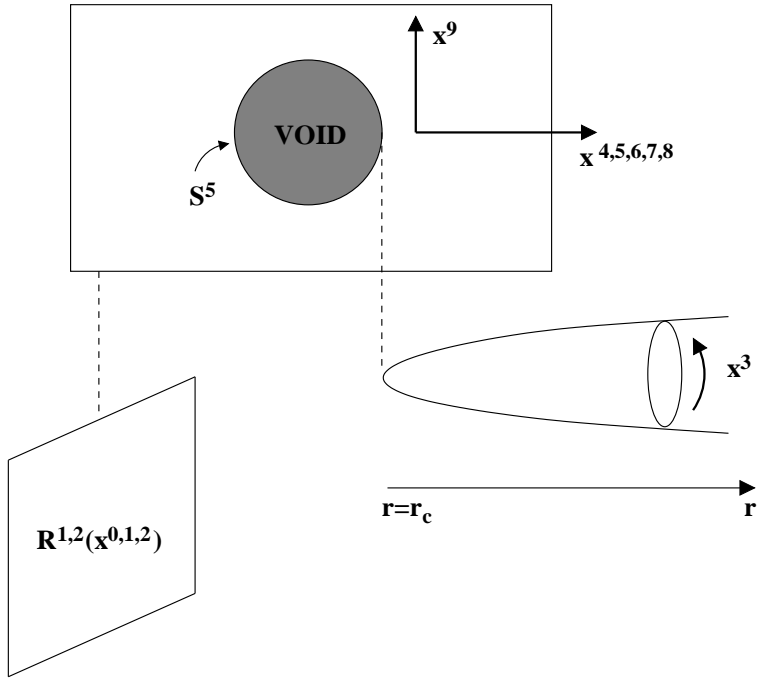}%
		\caption{\label{fig1}A schematic picture of the geometry produced by D3 branes.}
\end{figure}

To get a 3 dimensional $SU(N_c)$ Yang-Mills theory, one first considers $N_c$ D3 branes wrapping
a circle-compactified space direction, say $x^3\sim x^3+{2\pi \over M_{KK}}$, and spanning the flat non-compact 3 dimensional spacetime ${\mathbb R}^{1,2}=\{x^0,x^1,x^2\}$.  The world-volume theory on D3 branes is ${\mathcal N}=4$ super Yang-Mills theory, but
to break supersymmetry, one imposes by hand anti-periodic boundary condition on fermions in $x^3$ compactification. At sufficiently low energy and strong coupling, scalar fields are also expected to have loop-generated masses, and the theory flows to pure Yang-Mills theory at least in the sense of universality. According to the general philosophy of gauge/gravity correspondence, one then looks for
the corresponding back-reacted geometry produced by $N_c$ D3 branes by solving Type IIB supergravity equations of motion. The anti-periodic boundary condition for fermions is precisely parallel to
the Euclidean thermal circle compactification, and indeed this fact has been used to identify the solution
by suitable double-Wick rotations from a known black-brane solution at finite temperature.
The result is
\bear
ds^2&=&{r^2\over L^2}\left( f(r)\left(dx^3\right)^2+ \sum_{\mu=0,1,2}\left(dx^\mu\right)^2\right)
+{L^2\over r^2}{dr^2\over f(r)} +L^2 d\Omega_5^2\quad,\nonumber\\
F_5^{RR}&=& {\left(2\pi l_s\right)^4 N_c\over {\rm Vol}(S^5)}\epsilon_5\quad,\quad e^\phi=g_s\quad,
\eear
with
\be
L^4=4\pi g_s N_c l_s^4 \,\,,\quad f(r)=1-{M_{KK}^4 L^8\over 16}{1\over r^4}\,\,,\quad x^3\sim x^3+{2\pi\over M_{KK}}\,\,,
\ee
and $F_5^{RR}$ is normalized in our convention to satisfy
\be
{1\over (2\pi l_s)^4}\int_{S^5} \, F_5^{RR} = N_c\,\,.
\ee
The geometry is a warped product of a cigar-shaped $(r,x^3)$ part and a constant size $S^5$ as well as flat $(x^0,x^1,x^2)$ dimensions. The holographic radial dimension $r$ spans from the UV regime of $r=\infty$ down to a tip of the cigar at $r=r_c={M_{KK}L^2\over 2}$ beyond which geometry does not exist,  indicating a mass gap of the gauge theory due to confinement. A useful schematic picture of the geometry is shown in Fig.~\ref{fig1}.

Our interest is on adding fundamental flavor quarks in the model. We introduce  suitable flavor branes in the geometry,  neglecting their back-reactions {\it $\grave{a}$ la} Karch-Katz~\cite{Karch:2002sh}. Scanning through a few possibilities, we are lead to D7 branes sharing ${\mathbb R}^{1,2}=\{x^0,x^1,x^2\}$ with D3 branes, while additionally spanning five out of six transverse directions to the D3 brane world-volume \cite{rey}. Especially, this {\it excludes} the circle-compactified $x^3$ direction.
In the D-brane picture, this D3/D7 configuration is non-supersymmetric, because the number of string coordinates of Neumann and Dirichlet boundary conditions on each ends is $\sharp ND=6$. All massless D3-D7 strings are Ramond-sector fermions, which are precisely 3D Dirac fermions of fundamental representation under $SU(N_c)$. The number of D7 branes introduced is nothing but  the number of flavors ($N_F \ll N_c$) of 3D QCD. If one wishes, Dirac masses for flavor quarks can be introduced by moving  D7 branes away from the D3 branes along one common transverse direction of the 10 dimensional spacetime, which we will call $x^9$. Because 3D Dirac mass explicitly breaks parity and the parity-odd sign of the mass is precisely the sign of the D7/D3 separation in the $x^9$ direction, we see that the $x^9$ direction should change its sign  under the 3D parity operation.
As one of our motivations is to study parity anomaly of massless flavor quarks,  we will only focus on overlapping D3/D7 branes in the UV regime from now on.

Since the above consideration of D7 branes specifies only the boundary conditions in the UV regime, $r\to\infty$, the full embedding of D7 branes, extending to the deep IR region should be determined dynamically from
the world-volume dynamics of D7 branes in the given background geometry. One generically expects
non-vanishing dynamically-generated masses for fermions even with massless bare quarks in UV.
Recalling the parity-odd nature of $x^9$ direction as well as its interpretation as a UV bare mass,
we note that any bending of D7 branes in deep IR into a particular direction of $x^9$ should be interpreted as dynamical mass generation. Ref.~\cite{rey} first discussed this aspect in the current setting in some detail, though it focused on non-compactified D3 branes aiming at condensed matter applications.
Effects from the curved geometry in IR will be crucial for this, and this is the holographic manifestation of gauge dynamics on flavor quarks. In fact, quarks are tightly bound at strong coupling and form colorless mesonic degrees of freedom. The world-volume dynamics of $N_F$ flavor D7 branes is presumed to describe such flavor dynamics in our holographic description.

\begin{figure}[t]
	\centering
	\includegraphics[width=8cm]{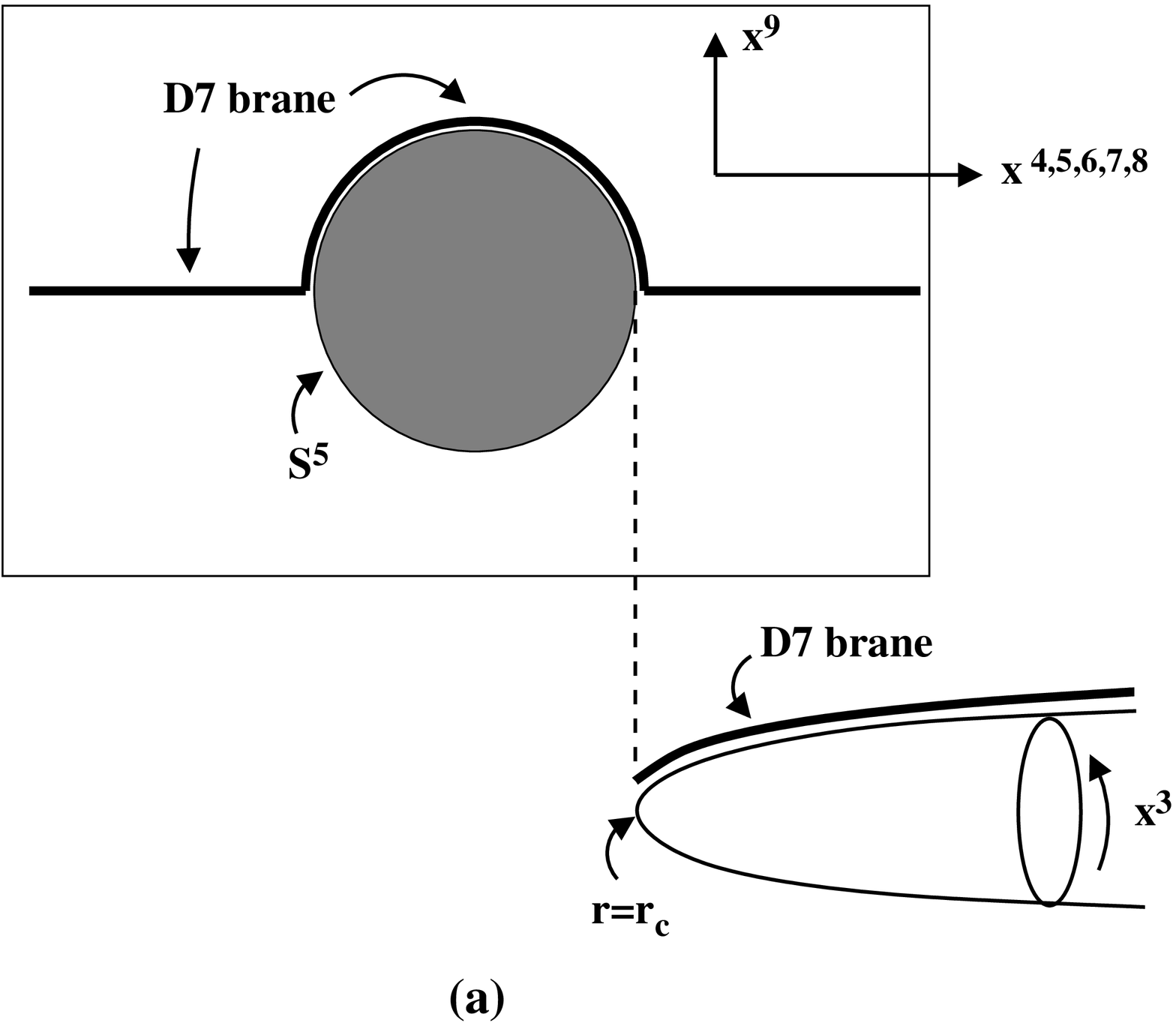}
 \includegraphics[width=8cm]{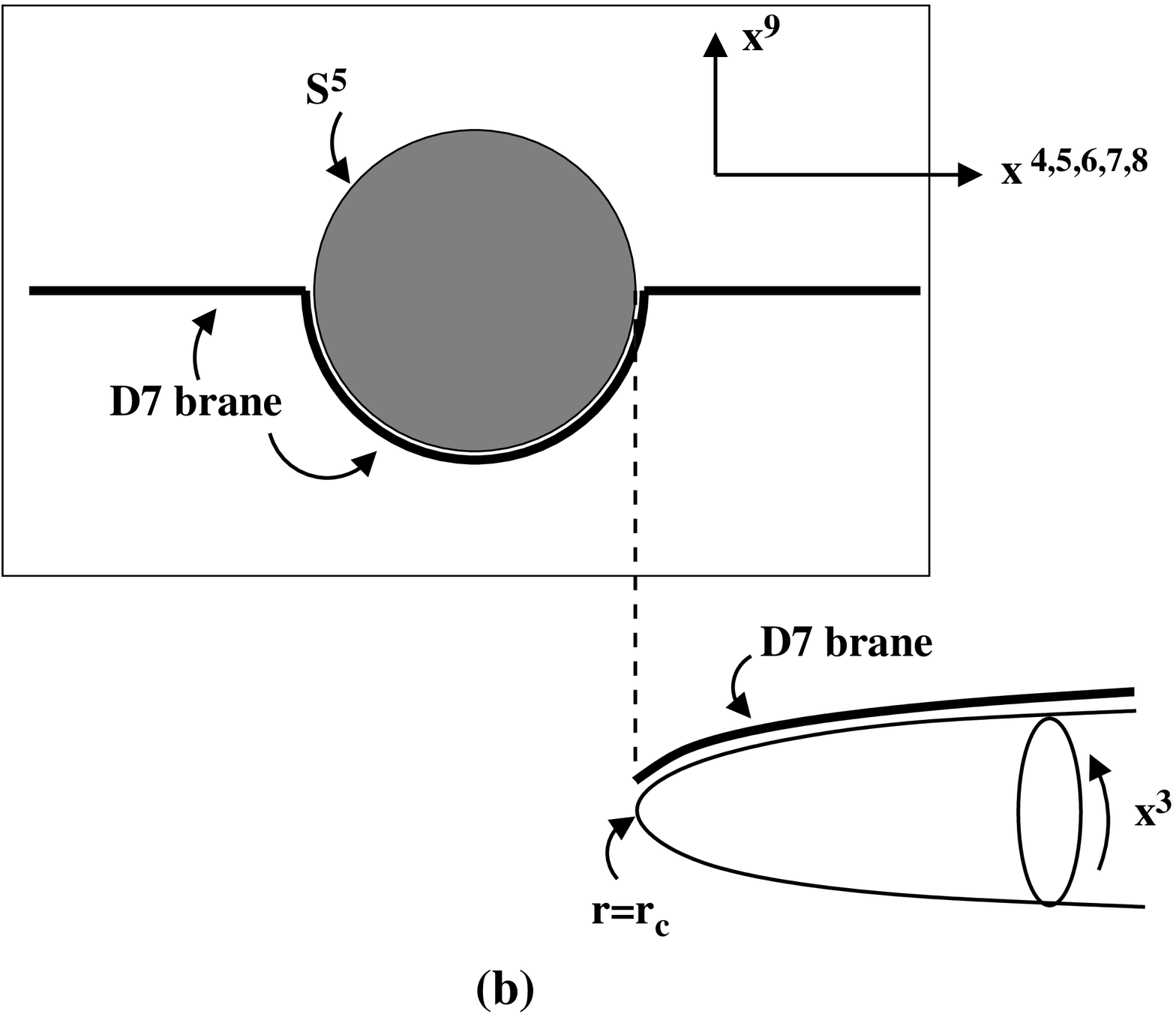}
		\caption{\label{fig2}Two possible D7 brane embeddings related by parity transformation.}
\end{figure}
\begin{figure}[t]
	\centering
	\includegraphics[width=8cm]{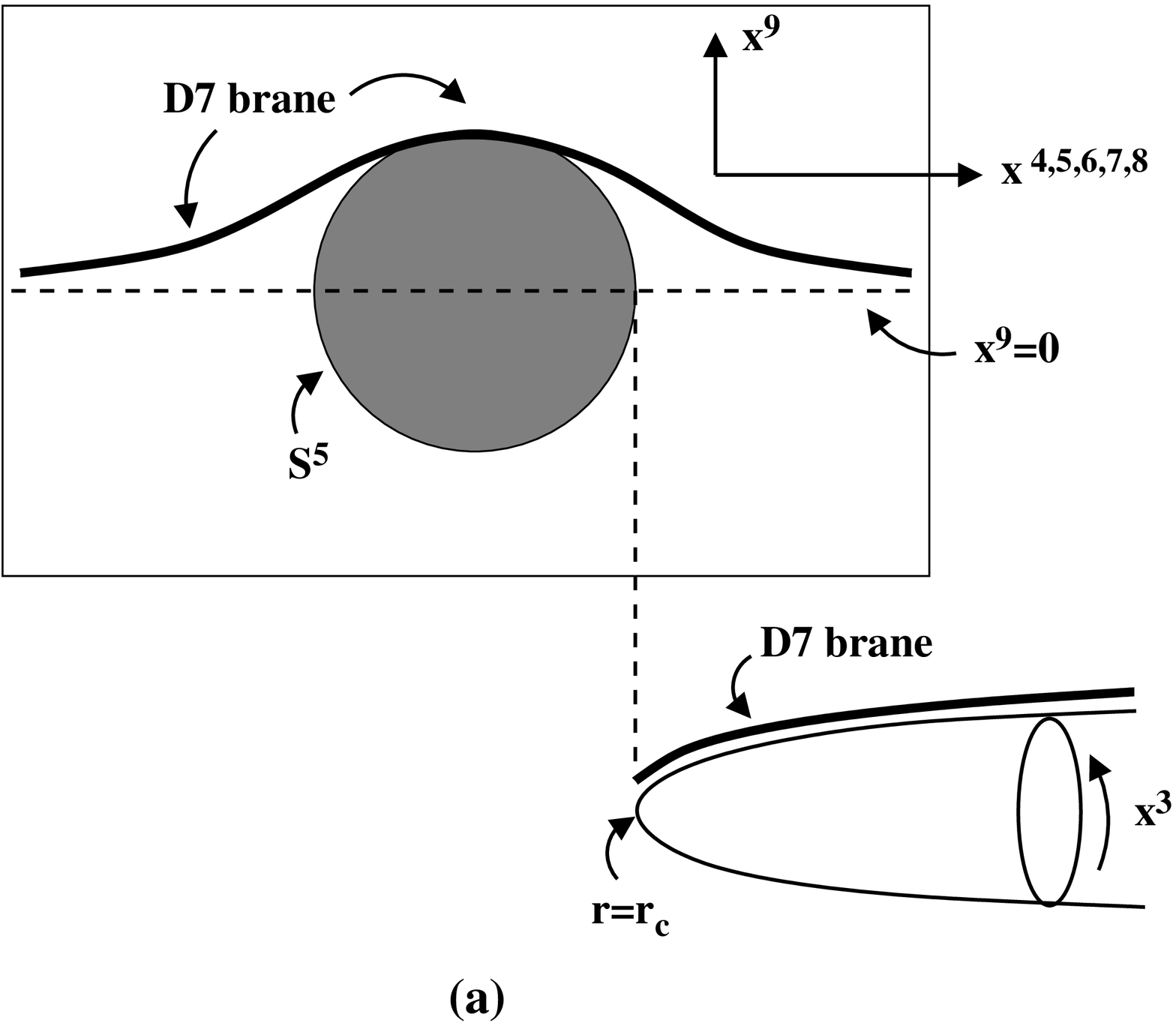}
 \includegraphics[width=8cm]{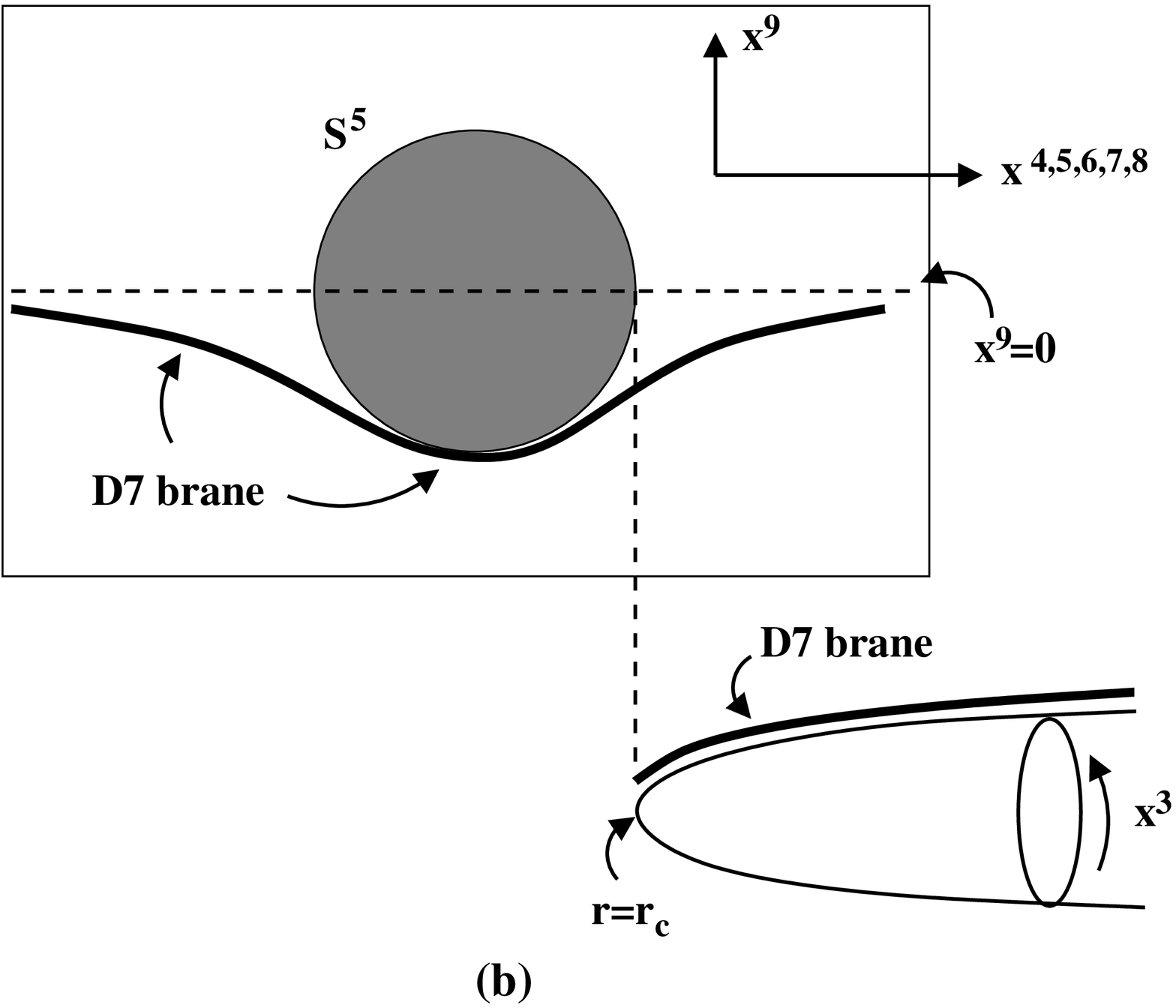}
		\caption{\label{fig3}More plausible D7 brane embeddings related by parity transformation.}
\end{figure}

In Fig.~\ref{fig2} we have tried to depict the holographic situation of a single ($N_F=1$) D7 brane embedding in the background produced by $N_c$ D3 branes. The UV boundary condition at $r\to\infty$ must be such as to ensure masslessness of bare quarks, and this simply means that the position of D7 branes in $x^9$ direction
should vanish as $r\to\infty$, if original D3 branes are sitting at $x^9=0$.
Because the geometry stops at $r=r_c$ with $S^5$ being the boundary, forgetting common ${\mathbb R}^{1,2}$ dimensions in our discussion, we see that there is no way for the D7 brane to stay at $x^9=0$ after reaching at the tip, otherwise its world-volume would have a boundary which is normally forbidden.
The way around is to wrap a half of $S^5$ to close its world-volume, and there are two possible ways, which are precisely related by the parity transformation of flipping $x^9$. Precisely speaking, the above configuration of D7 brane staying at $x^9=0$ until reaching at the tip and wrapping a half $S^5$ is not the one that minimizes its energy, and a more correct picture would be something like Fig.~\ref{fig3}.
One sees more clearly that the situation corresponds to  dynamical mass generation induced by geometry.
The statement of two possible embeddings which are related by parity operation still remains to hold  in Fig.~\ref{fig3}.

{\it Naively} the situation looks like {\it spontaneous breaking} of parity symmetry.
In UV region, parity seems unbroken as the $x^9$ position of the D7 brane embedding goes to zero.
Configuration (a) in Fig.~\ref{fig3} seems to be one ground state of the holographic 3D QCD with $N_F=1$, while configuration (b) is simply its parity partner state under parity operation, so one seems
to have just spontaneous breaking of parity in IR dynamics. However, this is {\it not} what we are after: parity anomaly is a statement of {\it unavoidable} parity breaking of the {\it theory} including its UV regime for massless 3D QCD, if gauge-invariance is maintained.

To clarify the situation more clearly, we take a short digression to explain one fact
that we will use which was originally discussed in Ref.~\cite{Fujita:2009kw}. Consider a completely different type of D7 brane introduced in our geometry of circle-compactified D3 branes: wrap a D7 brane over entire $S^5$ sitting at the tip $r=r_c$ while sharing the usual ${\mathbb R}^{1,2}$ dimensions.
Note that two dimensional transverse space to this D7 brane is nothing but the cigar of $(r,x^3)$,
where the $S^5$-wrapped D7 brane looks like a point source at the tip location. By a standard analogy
between D7 branes and two dimensional vortices, the D7 brane at the tip will introduce one unit of
$C_0^{RR}$ monodromy along $x^3$ circle, and this monodromy should extend even into the UV   regime at $r\to\infty$. This forces one to take the monodromy as a UV data in the definition of UV theory, rather than being a dynamically-generated IR phenomenon. In fact,
in the UV D-brane picture of $N_c$ D3 branes wrapping $x^3$ circle, one is allowed to introduce
$C_0^{RR}$ monodromy along compact $x^3$
\be
\int_{x^3} \, F_1^{RR} = k \quad,
\ee
and its effect on D3 brane world-volume can easily be deduced as
\be
\mu_3 {(2\pi l_s^2)^2\over 2!}\int_{D3} C_0^{RR}\wedge F\wedge F \sim  {1\over 4\pi}\int_{D3} F_1^{RR} \wedge A\wedge F = {k\over 4\pi}\int_{{\mathbb R}^{1,2}} A\wedge F\quad,
\ee
where $\mu_p=(2\pi)(2\pi l_s)^{-(p+1)}$ is the tension of p-brane, so that the effect is simply adding a Chern-Simons term in the 3D gauge theory Lagrangian. In the holographic picture at strong coupling, the monodromy requires $k$ number of D7 brane sources because the $x^3$ circle closes off at the tip and the geometry is topologically a disc rather than a cylinder. Energetically D7 branes would sit at the tip, and this is a more correct way of viewing a $S^5$-wrapped D7 brane at the tip.
In the weak-coupling D-brane picture, there is no obvious requirement of quantum values of $C_0^{RR}$ monodromy, while in the holographic strong-coupling picture, non-integer monodromy would be  inconsistent with the integer value of D7 brane number. This is therefore a holographic proof of the quantization of  Chern-Simons coefficients.
According to Ref.~\cite{Fujita:2009kw} this provides an interesting realization of level-rank duality of 3D gauge theory if one identifies the low energy theory on $k$ $S^5$-wrapped D7 branes as a 3D Chern-Simons theory of gauge group $U(k)$ of level $N_c$. For our purpose at present, it is enough to note that additional $S^5$-wrapped D7 brane corresponds to modifying the UV gauge theory by one unit of Chern-Simons term.
\begin{figure}[t]
	\centering
	\includegraphics[width=8cm,height=5.6cm]{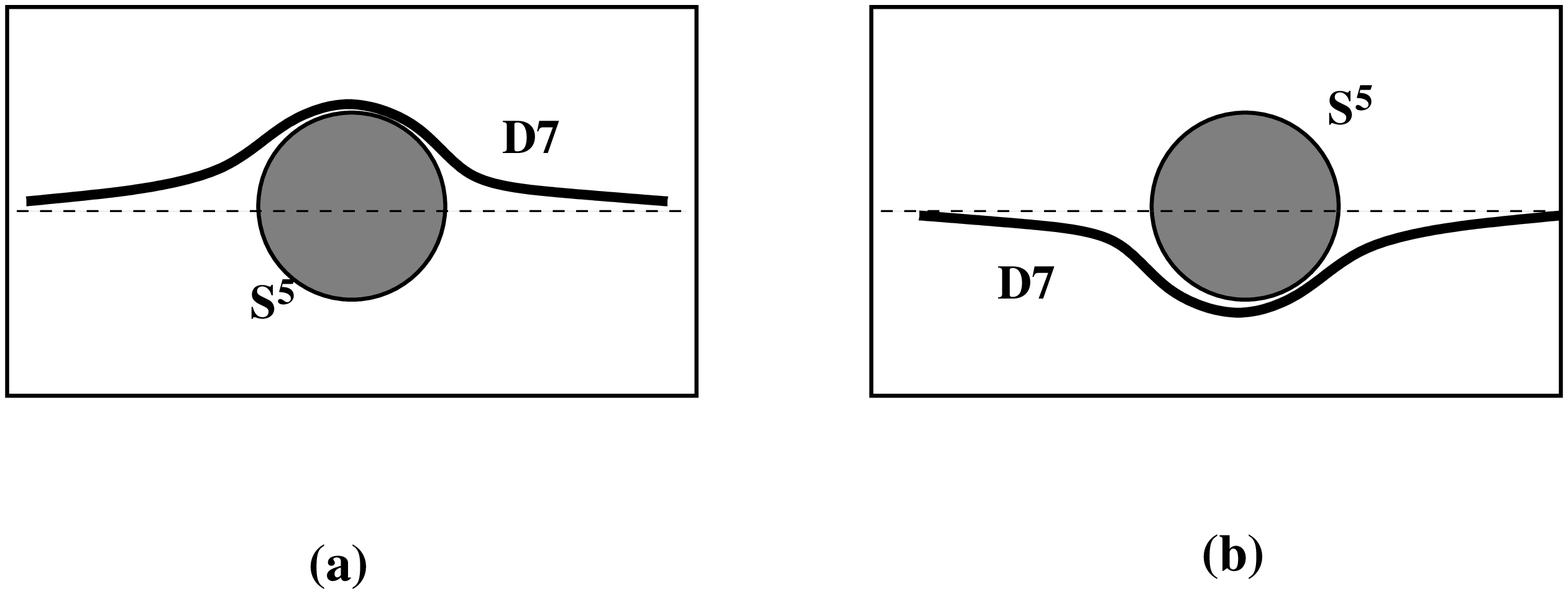}
 \includegraphics[width=8cm,height=5.3cm]{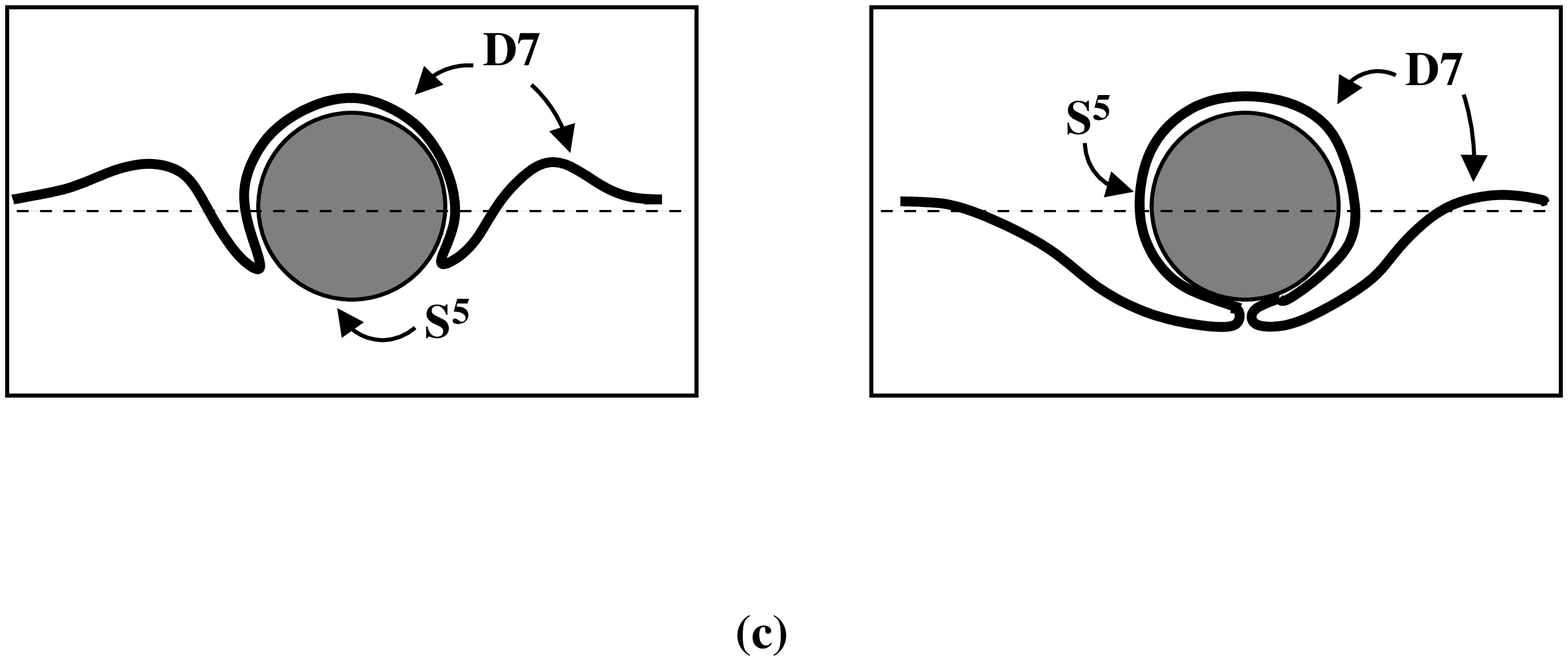}
		\caption{\label{fig4}The configuration (c) can be obtained from (a) by a smooth operation in IR regime, while the state (b) is disconnected from the states (a) and (c).}
\end{figure}

The above fact is important for our problem as can be seen in Fig.~\ref{fig4}. Starting from the state (a), one can smoothly deform the configuration only in the IR regime to reach the state (c). Because the operation involves only IR regime without modifyng the branes at UV region, the two states (a) and (c) can be considered as two different states in the {\it same} UV theory. The state (c) should be some meta-stable excited states above the ground state (a). Comparing (b) and (c), they differ from each other by a single $S^5$-wrapped D7 brane, and one concludes that the state (b) has a different UV theory from the one for (c) by one unit of Chern-Simons term in the gauge theory Lagrangian.
Therefore, {\it the states (a) and (b) do not belong to a same UV theory.} In other words, parity
transformation is not a symmetry of any UV theory of (a) nor (b) as it changes the theory from one to the other. There is no way to reinstate parity symmetry, and we argue this is a holographic manifestation of parity anomaly of $N_F=1$ 3D QCD.

A democratic treatment of the two UV theories corresponding to the states (a) and (b) would be that the theory of (a) has a {\it half} unit of Chern-Simons term explicitly while the theory of (b) has the negative one half, which can be thought of arising from UV regularizing such as Pauli-Villars.
Indeed, a single large mass Pauli-Villars fermion to regularize our dynamical fermion loop  gives us a half unit of Chern-Simons term contribution in the effective action whose sign depending on the choice of sign of the Pauli-Villars mass \cite{Redlich:1983kn}. As was observed originally by Redlich, this is consistent with large gauge transformations because the non-local finite piece in the total effective action from our massless dynamical fermion compensates the global gauge anomaly of the half unit Chern-Simons term, so that the total effective action is gauge invariant. What one looses here is the parity which is explicitly broken along the way by choosing the sign of Pauli-Villars mass. Indeed, the difference in the total effective action between two possible choices is a {\it single} unit of Chern-Simons term, which perfectly agrees with our holographic discussion.

\begin{figure}[t]
	\centering
	\includegraphics[width=5.3cm]{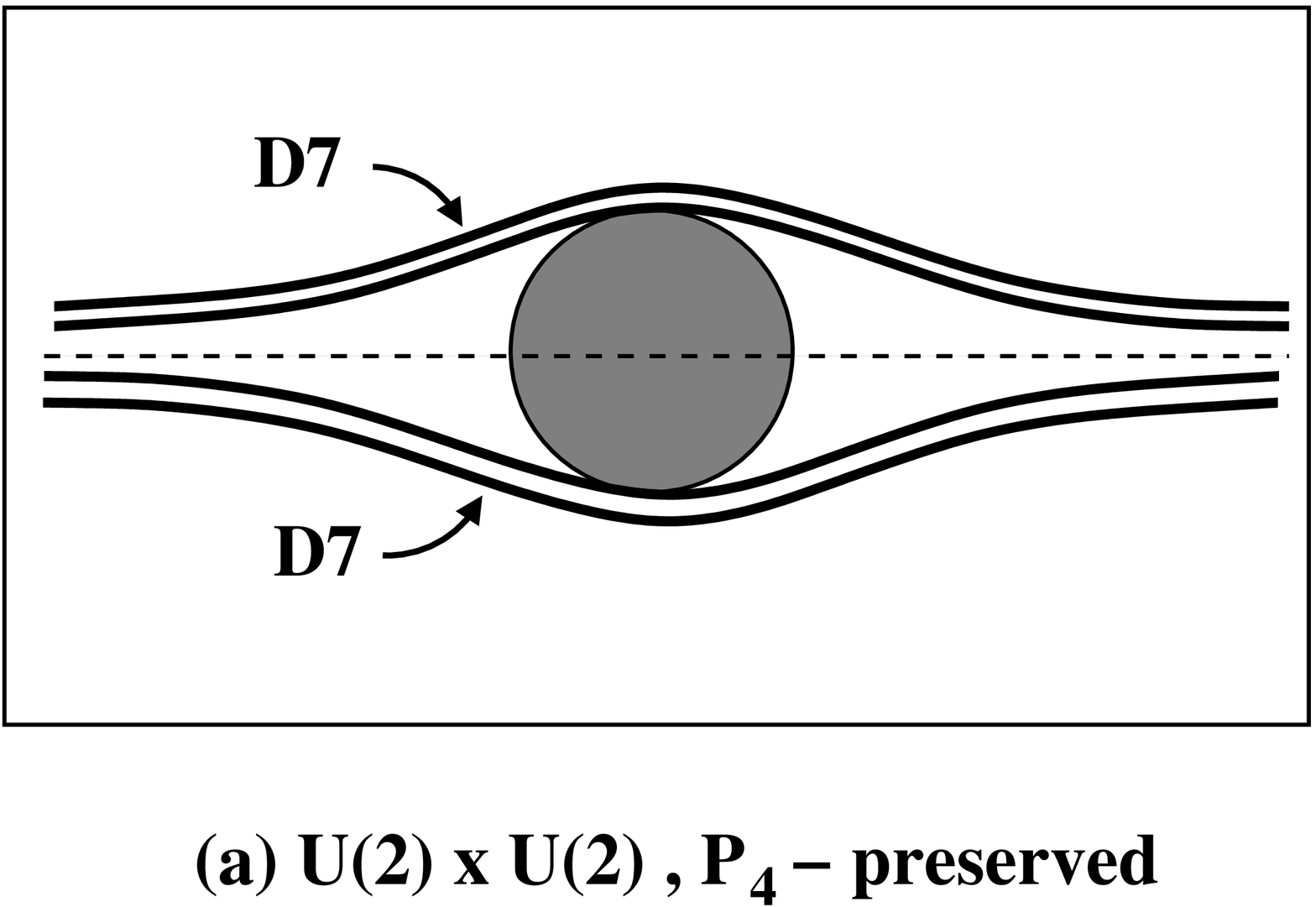}
 \includegraphics[width=5.3cm]{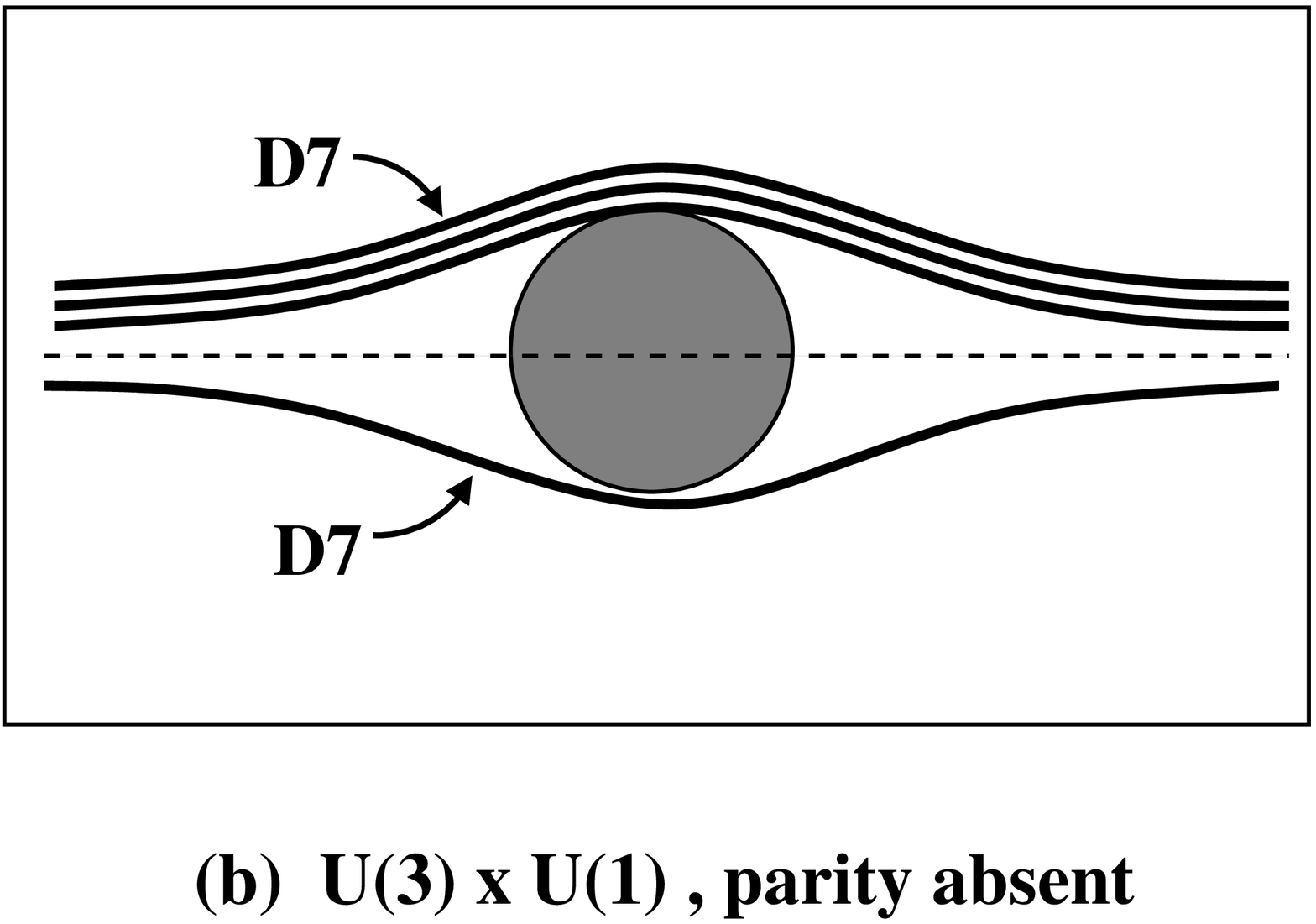}
 \includegraphics[width=5.3cm]{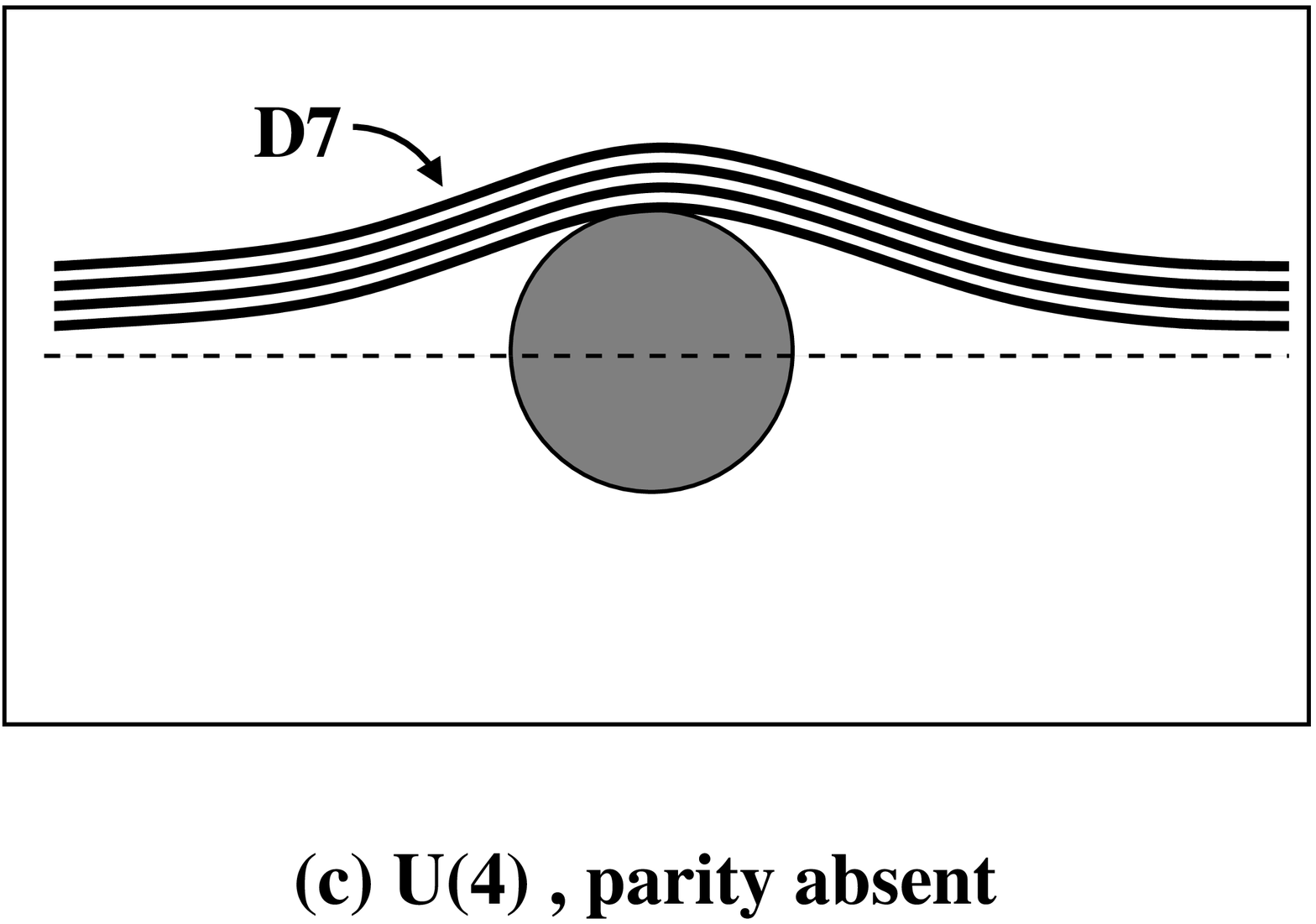}
		\caption{\label{fig5}Three theories with different IR flavor symmetry breaking patterns for $N_F=4$. All of them have $U(4)$ flavor symmetry at UV. The Vafa-Witten theorem applies only to (a) with $P_4$ parity symmetry.}
\end{figure}

Obviously, similar conclusion can be drawn
for any odd $N_F$, and for even $N_F$ one is {\it allowed} to have a configuration with equal number
of D7 branes on both sides of $x^9$ direction, and the UV theory for this configuration has a modified parity symmetry, called $P_4$ in our previous field theory discussion.
Note that in this case, the unbroken flavor symmetry group of the theory at UV is ${\rm U}(N_{F})$, though we have only ${\rm U}({N_F\over 2})\times {\rm U}({N_F\over 2})$ in the IR regime consistently with Vafa-Witten theorem.   {\it If one wishes} one can also have a theory of full $U(N_F)$ flavor symmetry in IR by placing all $N_F$ D7 branes in one side of $x^9$, but one has to give up parity to achieve this and Vafa-Witten theorem does not apply due to absence of parity symmetry.  However, the most interesting D7 brane configuration for 3D QCD is that of D7 branes preserving $P_{4}$. 
See Fig.~\ref{fig5} for an example of $N_F=4$.

\subsection{Weak-coupling D-brane picture}

\begin{figure}[t]
	\centering
	\includegraphics[width=10cm]{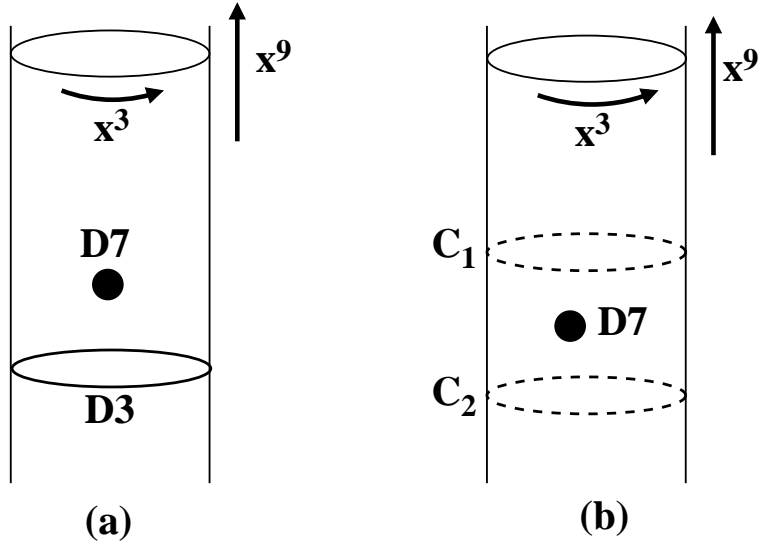}
		\caption{\label{fig6}A weak coupling D-brane picture of the D3/D7 brane system of our interest.}
\end{figure}

Because parity anomaly should be present even in the UV definition of the theory, we expect  it to exist in our specific UV completion of 3D QCD in terms of D3/D7 branes in weak-coupling limit. In this subsection we therefore try to give at least a reasonable account of this phenomenon in the D-brane picture.
The D-brane configuration is as follows: one has $N_c$ number of D3 branes spanning $x^{0,1,2,3}$ dimensions inside 10 dimensional space-time, and the $x^3$ direction is circle-compactified with radius $M_{KK}^{-1}$. The six transverse directions to the D3 brane world-volume are parameterized by $x^{4,5,6,7,8,9}$.
There are $N_F$ D7 branes sharing $x^{0,1,2}$ with the D3 branes and spanning five out of six transverse directions to the D3 branes, which we choose to be $x^{4,5,6,7,8}$. Therefore, the two dimensional transverse space to the D7 branes is $(x^3,x^9)$ which forms a cylinder. In our discussion, it will be enough to focus on this two-dimensional cylinder of $(x^3,x^9)$ on which D7 branes look like point sources, and D3 branes are 1-dimensional lines wrapping the circle direction $x^3$. Fig.~\ref{fig6}(a) is a schematic picture of this situation.
We will consider the case of $N_F=1$ for simplicity.

As discussed before, D7 brane is (magnetically) charged under $C_0^{RR}$ so that it causes a one unit of monodromy of $C_0^{RR}$ around a circle on its transverse space, in this case the cylinder.
Consider two contours $C_1$ and $C_2$ around $x^3$ on both sides of the D7 brane in $x^9$ direction as shown in Fig.~\ref{fig6}(b). Homologically, the difference $(C_1-C_2)$ is equivalent to a small circle contour around the D7 brane in the two dimensional space spanned by $x^{3}$ and $x^{9}$, so that there should be a one unit of $C_0^{RR}$ monodromy around it,
\be
\int_{C_1}\, F_1^{RR} -\int_{C_2}\, F_1^{RR} = 1\quad.\label{diff}
\ee
Imagine two situations of D3 branes sitting either side of $x^9$ with respect to the D7 brane. The D3 world-volume gauge theory contains a piece induced by background $C_0^{RR}$ field by
\be
{1\over 4\pi}\int_{D3}\, F_{1}^{RR}\wedge A\wedge F = {1\over 4\pi}\left[\int_{C_1 {\rm or} C_2}F_1^{RR}\right] \int_{{\mathbb R}^{1,2}} \, A\wedge F\quad,\label{d3top}
\ee
so that from (\ref{diff}) and (\ref{d3top}) one finds that the theory on D3 branes on one side of the D7 brane would differ from the one on the other side by one unit of Chern-Simons term. When we say massless fermion, it corresponds to D3 branes on top of the D7 brane, but this discussion indicates there is an intrinsic ambiguity in identifying correct $C_0^{RR}$ monodromy on the D3 brane world-volume to decide its Chern-Simons coefficient. This seems to be an explanation of parity anomaly in the D-brane picture. What really happens should be that there is a repulsion force between D3-D7 branes near intersection point which diverges at zero separation, so that D3 branes wave-function must {\it inevitably} bend to one side of the D7 brane near the intersection point microscopically. Diverging nature of the force would prohibit parity-symmetric configuration in any microscopic UV scale one desires, and one simply cannot have a parity-symmetric situation with any consistent UV complete D-brane picture.
The absence of supersymmetry of this D3/D7 brane configuration seems to support this expectation.

\section{Holographic Baby Skyrmions as Baryons}
\label{hbabyskyr}

In this section, we restrict our discussion to a parity-symmetric configuration with even number $N_F=2n$ of massless fundamental flavor quarks in the 3D QCD side, which corresponds to $N_F=2n$ D7 branes embedded symmetrically with respect to $x^9=0$ hypersurface in the gravity background generated by $N_c$ D3 branes. (so that there are $n$ D7 branes in each side of $x^9>0$ and $x^9<0$.)
The world-volume theory on the D7 branes is a gauge theory of $U(2n)$ flavor symmetry broken dynamically to $U(n)\times U(n)$ by separating D7 branes corresponding to a Higgs mechanism.
The expected bi-fermion condensate in 3D field theory side has a pattern
which preserves the flavor symmetry $U(n)\times U(n)$, and there is no spontaneous symmetry breaking of parity or vector symmetries.
This condensate is holographically encoded in the embedding shape of our D7 branes, but our discussion will be quite general so that we won't need details about the shape and the value of the condensate.
We are interested in how baryons are described in the set-up. See Refs.~\cite{Nawa:2006gv,Hong:2007kx,Hong:2007ay,Hata:2007mb,Park:2008sp,Hashimoto:2008zw,Grigoryan:2009pp} for baryons in the Sakai-Sugimoto model.

\begin{figure}[t]
	\centering
	\includegraphics[width=10cm]{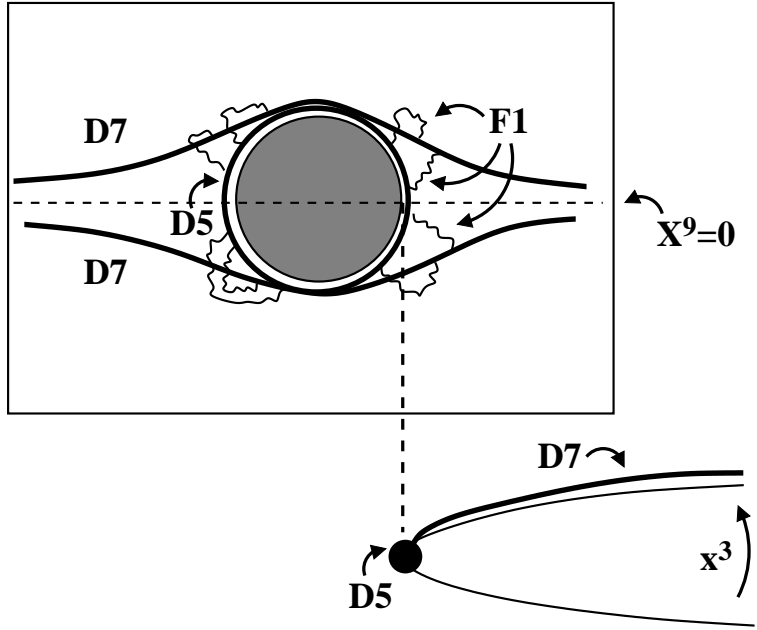}
		\caption{\label{fig7} A cartoon of $S^5$-wrapping D5 brane for baryon.}
\end{figure}
One can easily identify an object which carries baryon charge of $N_c$ number of fundamental quarks : a $S^5$ wrapped D5 brane which carries $N_c$ fundamental string charges due to the background $F_5^{RR}$ flux on $S^5$. The story is same as in the N=4 super Yang-Mills dual to $AdS_5\times S^5$ \cite{Witten:1998xy}.
What is special in our case is that we have D7 probe branes representing dynamical flavor quarks.
A first picture one can imagine is something like Fig.~\ref{fig7} where $N_c$ fundamental strings from the D5 brane are ending on the flavor D7 branes. Although this picture is perfectly ok, it is hard to say something more than the picture. If there is a complementary/alternative understanding on the same physics, it seems worthwhile to discuss it to deepen our intuition.

\begin{figure}[t]
	\centering
	\includegraphics[width=10cm]{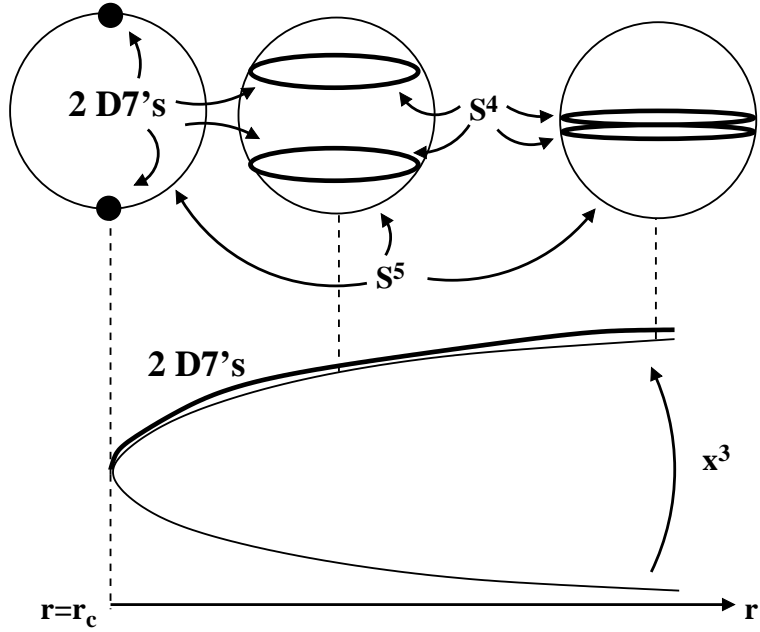}
		\caption{\label{fig8} Another way of viewing two D7 branes embedded in a parity-preserving manner.}
\end{figure}
Note that D7 branes are wrapping $S^4$ inside $S^5$, whose size depends on the position in the holographic coordinate $r$. Recalling that $S^5$ is a constant radius sphere in the space $x^{4,5,6,7,8,9}$ and $2n$ D7 branes are distributed symmetrically along $x^9=0$, we see that $n$ D7 branes are wrapping $S^4$ in one side of $x^9=0$ while $n$ others are wrapping the same size $S^4$ on the other side of $x^9=0$. $x^9=0$ defines a biggest $S^4$ inside $S^5$, and because we are considering massless bare quarks, each $n$ D7 branes should approach to this $S^4$ of $x^9=0$ in the UV boundary $r\to\infty$. See Fig.~\ref{fig8}. As D7 branes move towards IR region, their $S^4$ shrinks (in a symmetric way with respect to $x^9=0$) until
they reach a point $r=r_c$ where $S^4$ becomes a point and disappears (there are two points in either side of $x^9=0$ respectively). Note that D7 branes are sitting at a constant position in the compact $x^3$ coordinate in the cigar geometry of $(r,x^3)$ until they end at the tip $r=r_c$. Dynamically it is also possible to (and probably will) end before the tip, but our subsequent discussion will not be affected much by this. However, it can be easily checked that simple energetics would prohibit the D7 branes from going beyond $r=r_c$ and returning along the other side of $x^3$.

\begin{figure}[t]
	\centering
	\includegraphics[width=10cm]{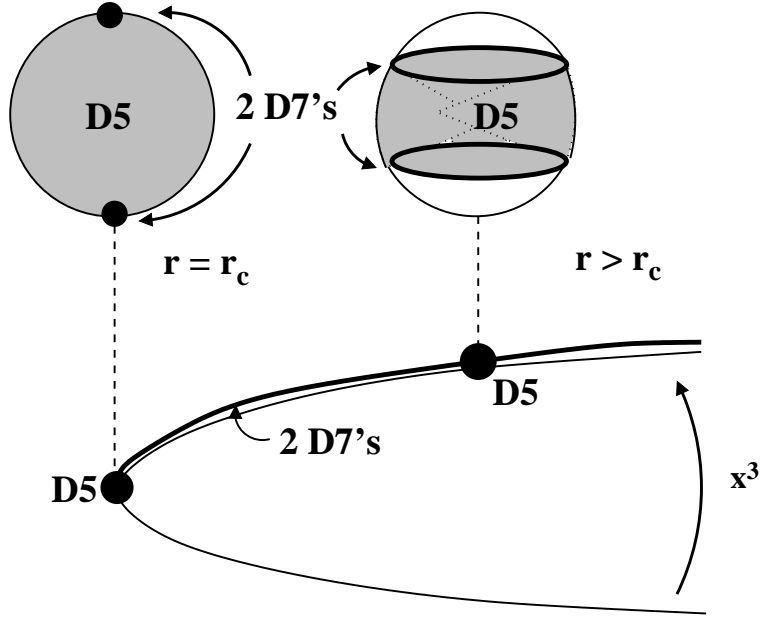}
		\caption{\label{fig9} D5 baryon at $r=r_c$ and $r>r_c$.}
\end{figure}

Consider a $S^5$ wrapped D5 brane as a baryon sitting at the IR tip $r=r_c$, where two sets of $n$ D7 branes on each sides of $x^9=0$ degenerate to points. These two degenerate points are intersections between the D5 brane and the two sets of $n$ D7 branes respectively. The important fact for us is that the D5 brane can now end on these two sets of D7 branes at the two intersection points, and its world-volume can be opened and suspended between the two sets of D7 branes as it moves towards $r>r_c$. Fig.~\ref{fig9} tries to depict the situation. This is a higher dimensional analogue of D3/D1 system where D1 brane can be suspended between two separated D3 branes, if one forgets common $S^4$ direction.
Curiously, the resulting D5 brane at $r>r_c$  {\it does not} wrap the full $S^5$, but only a part of it, so that the usual argument for its $N_c$ string charges coming from the background $F_5^{RR}$-flux does not work as it would give only a fraction of $N_c$ charge. However, the operation we have performed seems perfectly reasonable, especially it is an operation in the IR regime so that it should not have changed something like total charge of baryons dictated by symmetry. The purpose of this section is to give a reasonable answer to this puzzle.

Looking at the D3/D1 brane analogy further, an important fact is that D1 brane has an {\it alternative} D3 brane world-volume description in terms of 't Hooft-Polyakov monopole.
Therefore, the D5 brane suspended between two sets of D7 branes at $r>r_c$ can be identified as carrying a monopole charge $(+1,-1)$ with respect to $U(n)\times U(n)$ gauge symmetry on the D7 branes world-volume, where the charges sit in the trace part of $U(n)$.
Note that $S^4$ part in the spacetime is commonly occupied by D5/D7 branes, so that the monopole
will be present along the 4-dimensional holographic spacetime $(r,x^{0,1,2})$.
We correspondingly should consider the reduced world-volume gauge theory of D7 branes onto this 4-dimensional holographic spacetime by integrating over $S^4$ direction that they are wrapping.
Similarly to D3/D1 brane system, one expects that the $(+1,-1)$ monopole of $U(n)\times U(n)$ gauge theory on D7 branes reduced to $(r,x^{0,1,2})$ should be able to describe the original D5 brane baryon {\it at least qualitatively such as its baryon charge}. We point out that because of a small bulk string coupling constant, the size of this monopole is in fact a string-size, and a D7 field theory description like the one we propose here that necessarily neglects stringy effects is not a perfect one. However, what we would like to study is how the baryon charge in our picture can be understood, and because it is a symmetry-related question, one must be able to explain it even in the crude field theory set-up like ours. Eventually, our success that comes shortly will justify itself.

It is not difficult to find the $U(n)\times U(n)$ gauge theory in 4-dimensional spacetime $(r,x^{0,1,2})$ on the D7 branes after integrating over $S^4 \subset S^5$ that they are wrapping.
See Ref.\cite{Davis:2008nv} for a similar computation.
Because the monopole charge coming from a suspended D5 brane sits in the trace part of each $U(n)$'s, it is sufficient to consider the simplest case of $n=1$ or $N_F=2$.
As mentioned before, we don't need a specific detail about how D7 branes are embedded except the two boundary conditions at UV and IR : at UV boundary $r\to\infty$, the two D7 branes ($N_F=2$) wrapping two $S^4$'s on each side of the equator $x^9=0$ should approach to the common $S^4$ at the equator $x^9=0$.
At the IR end, the two $S^4$ on each side of $x^9=0$ degenerate to two points, say north/south poles,
and the D7 brane world-volumes end there. The common size of the two $S^4$'s is a monotonically increasing function of $r$. It is convenient to introduce a polar angle $0\le\theta\le\pi$ on $S^5$ so  that the $S^5$ metric can be written as
\be
d\Omega_5^2= d\theta^2 +\sin^2\theta d\Omega_4^2\quad,\label{s5metric}
\ee
where $d\Omega_4^2$ is a unit radius $S^4$ metric. $\theta=0$ ($\theta=\pi$) is the north (south) pole respectively, and $\theta={\pi\over 2}$ is the $x^9=0$ equator. Then D7 brane embedding is specified by the function $\theta(r)$. For the D7 brane in the upper hemisphere ($x^9>0$), $\theta(r)$ is a monotonic function from $\theta(r_c)=0$ to $\theta(r\to\infty)={\pi\over 2}$, while for the other D7 brane, $\theta(r)$ decreases monotonically from $\theta(r_c)=\pi$ to  $\theta(r\to\infty)={\pi\over 2}$.
Using this language, the induced metric on the D7 branes is given by
\be
g^*= {r^2\over L^2}\sum_{\mu=0,1,2} \left(dx^\mu\right)^2 +\left({L^2\over r^2 f(r)} +L^2 \left(d\theta\over dr\right)^2\right) dr^2 + L^2 \sin^2\theta(r) d\Omega_4^2\quad,
\ee
with $f(r)=1-{M_{KK}^4 L^8\over 16}{1\over r^4}$, and the world-volume action reads as
\be
S_{D7}=\mu_7\int d^8\xi\,\sqrt{{\rm det}(g^*+2\pi l_s^2 F)} +\mu_7 {(2\pi l_s^2)^2\over 2!}\int C^{RR}_4\wedge F \wedge F\quad,
\ee
where $\mu_7=(2\pi)^{-7}l_s^{-8}$. We assume homogeneity along $S^4$ direction and perform integration on it to get a 4-dimensional theory on $(r,x^{0,1,2})$. While the first term gives a DBI-type kinetic term for the world-volume gauge field, what is important for us is the second Chern-Simons term induced by the background RR field. Recall that $F_5^{RR}=dC_4^{RR}$ is normalized such that
\be
F_5^{RR} = {(2\pi l_s)^4 N_c\over {\rm Vol}(S^5)} \epsilon_5\quad,
\ee
where $\epsilon_5$ is the volume form of the unit radius $S^5$, and ${\rm Vol}(S^5)=\pi^3$ is its volume. Although $F_5^{RR}$ is well-defined, $C_4^{RR}$ is not, and it necessarily has a singularity
at a point on $S^5$. Writing $\epsilon_5=\sin^4\theta d\theta\wedge\epsilon_4$ with $\epsilon_4$ being the volume form of a unit $S^4$ (this can be inferred from the metric (\ref{s5metric})), a possible choice of $C_4^{RR}$ that preserves symmetries of $S^4$ is
\be
C_4^{RR}={(2\pi l_s)^4 N_c\over {\rm Vol}(S^5)}\left({3\over 8}\left(\theta-{\pi\over 2}\pm {\pi\over 2}\right) -{1\over 4}\sin 2\theta +{1\over 32}\sin 4\theta \right) \epsilon_4\quad,\label{c4}
\ee
where the upper(lower) sign case has a singularity at $\theta=\pi$ ($\theta=0$) respectively. When one integrates $C_4^{RR}$ over $S^4$ there are two different results depending on which choice one makes for $C_4^{RR}$. One may forgo the above explicit expression of $C_4^{RR}$ by simple use of Stokes formula
\be
\int_{S^4} \,C_4^{RR} = \int_{B^5}\,F^{RR}_5={(2\pi l_s)^4 N_c\over {\rm Vol}(S^5)} \cdot{\rm Vol}(B^5)\quad,
\ee
where $B^5$ is a ball on $S^5$ whose boundary is $S^4$, and obviously there are two possible choices for $B^5$ depending on where the singularity of $C_4^{RR}$ is chosen to be.

Not surprisingly by now, the choice of $C_4^{RR}$ is related to parity symmetry breaking : indeed, flipping of $x^9$ moves the singularity of one choice to the other, so that the resulting Chern-Simons term for {\it flavor symmetry} in the D7 brane world-volume action changes under parity transformation. Note that this is again an {\it explicit} breaking of parity symmetry in defining the theory, and it is another manifestation of the same parity anomaly.

To see this more clearly, focus on a single D7 brane whose embedding is in the upper hemisphere of $S^5$ ($x^9>0$). As $r$ goes from UV to IR, its world-volume spans the polar angle range of $0\le\theta\le{\pi\over 2}$, that is the whole upper hemisphere, so that one is in fact {\it forced}
to choose the singularity of $C_4^{RR}$ sitting at the lower hemisphere ($x^9<0$) in order to have a smooth world-volume action throughout spacetime. Therefore, this D7 brane must pick up the $C_4^{RR}$ in (\ref{c4}) with the upper sign. The resulting 4-dimensional term after integrating over $S^4$ would be then
\be
{\Theta(r)\over 8\pi^2} F\wedge F = {\Theta(r)\over 32\pi^2} \epsilon^{\alpha\beta\gamma\delta}F_{\alpha\beta} F_{\gamma\delta}\quad,
\ee
where the $r$-dependent $\Theta$ angle is given by
\be
\Theta(r)=\mu_7 {(2\pi l_s^2)^2\over 2!}{(2\pi l_s)^4 N_c\over {\rm Vol}(S^5)}{\rm Vol}(S^4)\left(8\pi^2 \right)F\left(\theta(r)\right)={16\over 3}N_c \,F\left(\theta(r)\right)\quad,
\ee
with
\be
F(\theta)=
\left({3\over 8}\theta -{1\over 4}\sin 2\theta +{1\over 32}\sin 4\theta \right)\quad,\ee
as in (\ref{c4}). From $\theta(r_c)=0$ and $\theta(\infty)={\pi\over 2}$, the $\Theta$ angle is a monotonic function of $r$ from zero at $r=r_c$ to $\Theta(\infty)=\pi N_c$ at the UV boundary.
Near the UV boundary one can approximately take a constant value of $\Theta=\pi N_c$, and this has a nice 3D gauge theory interpretation as follows. From $F\wedge F =d\left(A\wedge F\right)$, the 4-dimensional bulk $\Theta$ term can give rise to a boundary term at UV infinity
\be
{\Theta(\infty)\over 8\pi^2} A\wedge F ={N_c\over 8\pi} A\wedge F \Bigg|_{r=\infty}\quad,
\ee
and because $A(r=\infty)$ is interpreted as a non-dynamical potential that couples to the flavor symmetry current of the 3D QCD side, the above term is a Chern-Simons term for this {\it flavor symmetry} with level $k={N_c\over 2}$. Recall that in the previous section, we have seen that this D7 brane embedding
should correspond to the 3D gauge theory with an {\it explicit} parity-breaking Chern-Simons term of {\it SU($N_c$) gauge field} with level $k={1\over 2}$, which can be thought of arising from introducing gauge-invariant Pauli-Villars fermions to regularize divergence of fermion loops. Pleasingly this comes consistently with the above Chern-Simons term of flavor symmetry of level $k={N_c\over 2}$ because the same Pauli-Villars fermions of fundamental representation of $SU(N_c)$ will introduce a similar Chern-Simons term for flavor symmetry but with $N_c$ copies. The upshot is that the holographic set-up correctly reproduces {\it both gauge and flavor} Chern-Simons terms that are responsible for parity anomaly of $N_F=1$ 3D QCD. Ref.\cite{Davis:2008nv} obtained a similar Chern-Simons term for flavor symmetry part in a slightly different non-compact D3/D7 set-up.
Note that these are just UV boundary terms induced by regularization such as Pauli-Villars fermions, and the {\it full} effective action after integrating over all bulk fields should contain  {full integral} Chern-Simons term $k=N_c$ as in (\ref{eff}) for consistency. To check this expectation more rigorously, one needs to perform careful holographic renormalization, which is left for a future work.

Back to our parity-symmetric case of $N_F=2$, one easily sees that the other D7 brane embedded in the lower hemisphere of $S^5$ ($x^9<0$) has to choose the other $C_4^{RR}$ of singularity sitting at the upper hemisphere and the resulting  4-dimensional $\Theta$ angle is precisely negative to the one we presented before for the D7 brane which wraps the upper hemisphere. The $U(1)\times U(1)$ gauge theory thus has the $\Theta$ term of structure $(\Theta(r),-\Theta(r))$.
Although this term might seem to break $U(2)$ symmetry explicitly down to $U(1)\times U(1)$ even at UV scale, it has no dynamical consequences at UV when we turn off external boundary flavor gauge fields couple to flavor currents, as discussed previously in the field theory viewpoint. This is because bulk $\Theta$-term is a total divergence around UV regime where $\Theta$ becomes constant, and its UV boundary integral vanishes in the absence of external flavor gauge fields. Its effects in the IR regime should be considered as due to dynamical mass generations, not due to any explicit UV breaking. It is precisely in this sense that one can say {\it spontaneous breaking} of $U(2)$ to $U(1)\times U(1)$ dynamically.

Be reminded of that our purpose is to study a monopole of charge $(+1,-1)$ in this 4-dimensional holographic gauge theory, especially to explain how the expected electric charge of $N_c$ under the diagonal $U(1)$ can be understood.
One is immediately lead to the famous Witten effect stating that a unit monopole in the presence of non-vanishing $\Theta$ becomes in fact a dyon of electric charge $Q_e=-{\Theta\over 2\pi}$ \cite{Witten:1979ey}. This additional electric charge should be considered as a {\it shift} of the original electric charge of the particle. In other words, if the original particle is a dyon of {\it bare} electric charge $Q_b$ with unit monopole strength, then the effect of $\Theta$-angle is to shift the electric charge to $\left(Q_b-{\Theta\over 2\pi}\right)$. In our case, we have a monopole charge of $(+1,-1)$ with $(\Theta(r),-\Theta(r))$, so the total shift is $-{\Theta(r)\over\pi}$. In fact, it is natural to expect that our object originally has an electric charge $N_c$ from the start. This is clear when the D5 brane is sitting at the IR boundary $r=r_c$ where it wraps the whole $S^5$ and has the charge $N_c$. The $\Theta$-angle vanishes at this point $r=r_c$, so there is no ambiguity about its total charge. When it moves to $r>r_c$, the $\Theta$-angle is no longer zero, and we need to understand the total electric charge carefully, which is the purpose of this section. Because the process is smooth, it is natural to expect that the original {\it bare} electric configuration of charge $N_c$ should not change, while there seems to be a shift in {\it physical} electric charge due the Witten effect. Moreover, our problem is more puzzling because $\Theta$ angle is $r$-dependent taking values of zero at $r=r_c$ and $\pm \pi N_c$ at $r=\infty$, so that the physical electric charge not only differs from $N_c$, but also seems dependent on the position where it sits. In fact, one can easily check that this {\it naive} physical electric charge $\left(N_c-{\Theta(r)\over \pi}\right)$ is precisely the value one obtains from the portion of $S^5$ that the D5 brane is wrapping through the $F_5^{RR}$-flux as in Fig.\ref{fig9}.

To account for the missing electric charge ${\Theta(r)\over \pi}$ properly,
we need to treat
the physics of space-varying $\Theta$ angle more carefully, which was first studied by Lee \cite{Lee:1986mm}. A general lesson from his work is that in addition to the electric charge shift from local Witten effect, there are other electric charges deposited in the surrounding medium proportional to the gradient of $\Theta$ angle, so that the {\it total} electric charge of the system can be different if $\Theta$ angle is not a constant. It is this effect that solves our puzzle as we show shortly : we will find that the medium-deposited electric charge due to the $r$-dependent $\Theta$-angle precisely explains the missing electric charge ${\Theta(r)\over \pi}$, so that the total physical charge in fact remains to be $N_c$.

To make the discussion general, let us consider a single $U(1)$ theory with space-varying coupling constant and $\Theta$ angle,
\be
{\cal L}_{4D} = -{1\over 4 e^2} F^2+ {\Theta\over 32\pi^2}\epsilon^{\alpha\beta\gamma\delta} F_{\alpha\beta} F_{\gamma\delta}\quad,
\ee
where space-dependence of $e^2$ and $\Theta$ is assumed, and we take a simple form of kinetic term in flat metric instead of the DBI-like expression in our original problem to simplify the discussion, because our focus is on how electric charges appear and the details of kinetic term are not important as the mechanism is topological in nature.
The conjugate momenta of the dynamical variables $A_i$ ($i=1,2,3$) are
\be
\Pi_i = {\delta {\cal L}_{4D}\over \delta\left(\partial_0 A_i\right)} = {1\over e^2} F_{0i}+{\Theta\over 8\pi^2}\epsilon^{ijk}F_{jk}\equiv{1\over e^2} E_i + {\Theta\over 4\pi^2} B_i\quad,
\ee
and it is natural to have commutation relations
\be
\left[A_i(x),\Pi_j(y)\right]=i \delta_{ij} \delta^{(3)}(x-y)\quad,
\ee
so that $\Pi_i \sim -i{\delta \over \delta A_i}$ acting on quantum wave-functionals of $A_i$.
In the usual temporal gauge $A_0=0$, one imposes the equation of motion of $A_0$ as a constraint on the physical Hilbert space, which is called the Gauss's law.
Computing the equation of motion of $A_0$, one finds
\be
\partial_i \Pi_i =0\quad,
\ee
which is equivalent to the invariance of wave-functionals under gauge transformations $A_i\to A_i+\partial_i \alpha$. In the presence of static external sources of electric charges $\rho_e$, the Gauss constraint is modified as
\be
\partial_i \Pi_i = -\rho_e\quad,
\ee
which can be seen most easily by considering a term $A_0\rho_e$ in the Lagrangian.
Therefore, it is $-\partial_i \Pi_i$ that measures {\it physical} electric charge density.
The Hamiltonian density is easily computed as
\be
{\cal H}=\left(\partial_0 A_i\right)\Pi_i -{\cal L}_{4D} = {e^2\over 2}\left(\Pi_i\Pi_i
-{\Theta\over 2\pi^2}\Pi_i B_i \right)+{1\over 2 e^2}\left(1+\left(e^2\Theta\over 4\pi^2\right)^2\right) B_i B_i\quad,
\ee
and from this and the commutation relation, the equations of motion in the Heisenberg picture are
\bear
 {d A_i\over dt} &=& E_i =i\left[\int d^3x\,{\cal H},A_i\right]={\partial{\cal H}\over \partial\Pi_i}
=e^2\left(\Pi_i -{\Theta\over 4\pi^2}B_i\right)\quad,\nonumber\\
{d\Pi_i\over dt} &=& i\left[\int d^3x\,{\cal H},\Pi_i\right]=-\epsilon^{ijk}\partial_j\left({1\over e^2}\left(1+\left(e^2\Theta\over 4\pi^2\right)^2\right)B_k\right)\quad.\label{hamiltoneq}
\eear
The first equation is a trivial identity while the second equation is a quantum version of the classical Maxwell equation. Let us study this theory semiclassically, meaning that we consider semiclassical expectation values of operators which satisfy the above equations of motion.
Then consider a particle with {\it bare} electric configuration of charge $Q_b$ and unit magnetic charge sitting at a point $\vec x_0$ relevant for our problem.
By definition, we have a {\it static} configuration satisfying
\be
\partial_i \left(E_i\over e^2\right)=- Q_b \delta^{(3)}(\vec x -\vec x_0)\quad,\quad \partial_i B_i = 2\pi \delta^{(3)}(\vec x -\vec x_0)\quad.\label{source}
\ee
The solution of the problem proceeds in two steps.
The second equation in (\ref{hamiltoneq}) and  (\ref{source}) give
\be
\vec \nabla \times \left({1\over e^2}\left(1+\left(e^2\Theta\over 4\pi^2\right)^2\right)\vec B\right)=0\quad,\quad \vec \nabla\cdot\vec B = 2\pi \delta^{(3)}(\vec x -\vec x_0)\quad,\label{beq}
\ee
from which one finds a solution for $\vec B$ using techniques of classical electrostatics. Then, the first equation of (\ref{hamiltoneq}) gives
\be
\vec \Pi = {1\over e^2} \vec E+ {\Theta\over 4\pi^2}\vec B\quad,
\ee
and one can compute the physical charge density of the system by
\be
\rho_e= -\vec\nabla\cdot\vec\Pi = -\vec\nabla\cdot\left({1\over e^2} \vec E+{\Theta\over 4\pi^2}\vec B\right)=-{1\over 4\pi^2}(\vec\nabla \Theta)\cdot \vec B+\left(Q_b-{\Theta(\vec x_0)\over 2\pi}\right)\delta^{(3)}(\vec x -\vec x_0)\quad,\label{eden}
\ee
where the last term is identified as the famous local Witten effect and the first term represents electric charges deposited in the medium proportional to the $\Theta$-gradient, which is what we are looking for.
Although the equation (\ref{beq}) that dynamically determines profiles of $\vec B$ will change as one uses DBI-like kinetic term in a curved metric instead of our simplified kinetic term, the end result for the electric charge density (\ref{eden}) can  easily be checked to hold universally.
One sees that electric charges in the medium depend on the profile of the magnetic flux $\vec B$, as well as the $\Theta$-angle profile. In our situation, the total medium-deposited electric charge density is a twice of the first term in (\ref{eden}) due to two copies of $U(1)$'s :
\be
\Delta\rho=-{1\over 2\pi^2}(\vec\nabla \Theta)\cdot \vec B\quad.\label{meddep}
\ee

\begin{figure}[t]
	\centering
	\includegraphics[width=8cm]{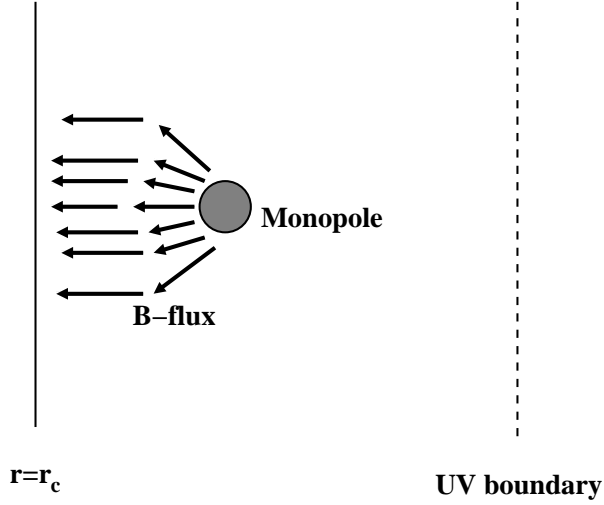}
		\caption{\label{fig10} An expected behavior of $\vec B$ out of monopoles.}
\end{figure}

Although the details of the shape of the magnetic flux $\vec B$ out of the object
can only be found by solving an analog of (\ref{beq}) in the full DBI action, it is not difficult to find the total integrated medium-deposited charge from (\ref{meddep}) from a  reasonably assumed global feature of the magnetic field $\vec B$, which is schematically depicted in Fig.~\ref{fig10}.
Because the energy cost for having a magnetic flux decreases as one goes to more IR region, the unit magnetic flux coming from the object will eventually head towards the IR boundary at $r=r_c$, where the world-volume of the D7 branes ceases to exist. Note that $\vec B$ along the radial direction $B_r$ is nothing but the component $F_{12}$, and there is no singular behavior associated to the above picture of magnetic flux ending at $r=r_c$, because $r=r_c$ is a smooth end of the D7 world-volume as the wrapped $S^4$ shrinks to zero.
Then the integrated total medium-deposited electric charge is
\bear
\Delta Q&=& \int d^2x \int_{r_c}^r dr' \,\Delta\rho(r') = -{1\over 2\pi^2}\left[\int d^2 x\, B_r\right]
\int_{r_c}^r dr'\, \partial_{r'} \Theta(r')\nonumber\\ &=& -{1\over 2\pi^2}\cdot\left(-2\pi\right)\cdot\left(\Theta(r)-\Theta(r_c)\right) = {\Theta(r)\over \pi}\quad,
\eear
where we used the unit $2\pi$ total magnetic flux heading towards $r=r_c$ at each fixed value of $r'$,
and also $\Theta(r_c)=0$. The result is precisely what one needs to make the total electric charge
of the system, that is the sum of the dyonic charge shifted by Witten effect and the charge deposited
in the medium, to be $N_c$.

\section{Discussion}

 In this paper we have considered a specific D3/D7 brane set-up for 3D QCD with $N_F$ number of massless quarks. The strongly-coupled IR regime of 3D QCD is then analyzed  in the supergravity limit or large $N_c$ limit, which is shown to realize geometrically the spontaneous breaking of ${\rm U}(N_F)$ ``chiral" symmetry down to the parity-invaraint
${\rm U}\left(\frac{N_F}{2}\right)\times{\rm U}\left(\frac{N_F}{2}\right)$, when $N_F$ is even.
The holographic aspect of parity anomaly in 3D QCD is studied in detail to show that it correctly leads to  Chern-Simons terms for both color gauge and flavor symmetry.

We have found that the holographic dual of 3D QCD at low energy  is a ${\rm U}(N_F)$ gauge theory dynamically broken down to ${\rm U}({N_F\over 2})\times {\rm U}({N_F\over 2})$ with  a parity-invariant $\theta$ term in four dimensions of nontrivial supergravity background. We then showed that the baryons or baby Skyrmions of low-energy QCD in three dimensions are realized as magnetic monopoles (more precisely, dyons of unit magnetic monopole charge) of the holographic dual theory. The baryon number of the objects can be explained from the Witten mechanism plus a novel effect from the radial dependent $\Theta$-angle. We expect the size of baryons to be stringy, and it needs further study to find their properties in detail, which is left for the future.

The low energy dynamics of three dimensional QCD shares many properties with strongly-coupled planar condensed matter systems, exhibiting same symmetry breaking pattern. So, they might belong to same universal classes. Unlike strongly-coupled planar condensed matter, however, 3D QCD has a natural holographic picture from string theory, based on D3/D7 branes.  It therefore might shed light on finding gravity dual theories of planar condensed-matter systems \cite{rey,Rey:2008zz}.
It will be
interesting to see if the spectrum of
baryon monopoles exhibits strange metallic behavior, found recently in~\cite{Lee:2008xf,Liu:2009dm,Cubrovic:2009ye,Faulkner:2009wj,Hartnoll:2009ns},  when the baryon chemical potential is turned on.

\eject
\acknowledgments
D.K.H. thanks S. A. Hartnoll and P. Yi for useful discussions.
The work of D.K.H. was supported by the Korea Research
Foundation Grant funded by the Korean Government (KRF-2008-341-C00008). Part  of this work was done during APCTP focus program FP-02, 2009 and also at the Aspen Center for Physics, whose stimulating environment is greatly acknowledged. H.U.Y. thanks organizers of the focus program for hospitality.

\end{document}